\documentclass[pdflatex,sn-mathphys]{sn-jnl}

\jyear{2023}%

\theoremstyle{thmstyleone}%
\theoremstyle{thmstyletwo}%

\theoremstyle{thmstylethree}%

\usepackage{amssymb}

\usepackage{bbm}
\usepackage{array}
\usepackage{mdwtab}
\usepackage{eqparbox}
\usepackage{makecell}
\usepackage{graphicx}
\usepackage{booktabs, multirow} 
\usepackage{soul}
\usepackage{changepage,threeparttable} 
\usepackage{rotating}
\usepackage{tabularx} 

\begin{document}

\title[IntegrAO]{Integrate Any Omics: Towards genome-wide data integration for patient stratification}




\author[1,2,3]{\fnm{Shihao} \sur{Ma}}
\author[5,8]{\fnm{Andy G.X.} \sur{Zeng}}
\author[2,6,8]{\fnm{Benjamin} \sur{ Haibe-Kains}}
\author[2,3,7]{\fnm{Anna} \sur{Goldenberg}}
\author[5,8]{\fnm{John E} \sur{Dick}}
\author*[1,2,3,4]{\fnm{Bo} \sur{Wang}}\email{bowang@vectorinstitute.ai}

\affil[1]{\orgdiv{Peter Munk Cardiac Centre}, \orgname{University Health Network}, \orgaddress{\city{Toronto}, \state{Ontario}, \country{Canada}}}

\affil[2]{\orgdiv{Vector Institute for Artificial Intelligence}, \orgname{}, \orgaddress{\city{Toronto}, \state{Ontario}, \country{Canada}}}

\affil[3]{\orgdiv{Department of Computer Science}, \orgname{University of Toronto}, \orgaddress{\city{Toronto}, \state{Ontario}, \country{Canada}}}

\affil[4]{\orgdiv{Department of Laboratory Medicine and Pathobiology}, \orgname{University of Toronto}, \orgaddress{\city{Toronto}, \state{Ontario}, \country{Canada}}}

\affil[5]{\orgdiv{Department of Molecular Genetics}, \orgname{University of Toronto}, \orgaddress{\city{Toronto}, \state{Ontario}, \country{Canada}}}

\affil[6]{\orgdiv{Department of Medical Biophysics}, \orgname{University of Toronto}, \orgaddress{\city{Toronto}, \state{Ontario}, \country{Canada}}}

\affil[7]{\orgdiv{Genetics and Genome Biology}, \orgname{the Hospital for Sick Children}, \orgaddress{\city{Toronto}, \state{Ontario}, \country{Canada}}}

\affil[8]{\orgdiv{Princess Margaret Cancer Centre}, \orgname{University Health Network}, \orgaddress{\city{Toronto}, \state{Ontario}, \country{Canada}}}

\abstract{

High-throughput omics profiling advancements have greatly enhanced cancer patient stratification. However, incomplete data in multi-omics integration presents a significant challenge, as traditional methods like sample exclusion or imputation often compromise biological diversity and dependencies. Furthermore, the critical task of accurately classifying new patients with partial omics data into existing subtypes is commonly overlooked. To address these issues, we introduce  \textbf{IntegrAO} (\textbf{Integr}ate \textbf{A}ny \textbf{O}mics), an unsupervised framework for integrating incomplete multi-omics data and classifying new samples. IntegrAO first combines partially overlapping patient graphs from diverse omics sources and utilizes graph neural networks to produce unified patient embeddings. Our systematic evaluation across five cancer cohorts involving six omics modalities demonstrates IntegrAO's robustness to missing data and its accuracy in classifying new samples with partial profiles. An acute myeloid leukemia case study further validates its capability to uncover biological and clinical heterogeneity in incomplete datasets. IntegrAO's ability to handle heterogeneous and incomplete data makes it an essential tool for precision oncology, offering a holistic approach to patient characterization. }


\maketitle

\section{Introduction}\label{sec1}


Precision medicine, which tailors personalized treatment based on the unique genetic profiles of individual cancer patients, has been recognized as the foundation of future cancer therapeutics \cite{shin2017precision}. The field is moving towards gathering multimodal data to address cancer's inherent heterogeneity \cite{steyaert2023multimodal}, characterized by diverse genetic, transcriptomic, and phenotypic variations \cite{belizario2019insights, lynch2015milestones}. Recent advancements in high-throughput technologies have enabled multi-dimensional profiling through diverse omics modalities. Projects like The Cancer Genome Atlas (TCGA) \cite{cancer2008comprehensive} and the International Cancer Genome Consortium (ICGC)\cite{zhang2011international} have produced and collected thousands of tumor samples at different molecular levels. Moreover, the rise of single-cell profiling, particularly single-cell transcriptomics, has deepened insights into tumor microenvironments by highlighting the distinct expression profiles of various cell types. Consequently, patient stratification, which involves categorizing patients based on distinct genetic, transcriptomic, and phenotypic profiles, has become a critical process in precision medicine for aiding in the development of tailored treatment approaches.


Integrating multi-omics data, leveraging the complementary nature of these datasets, offers a more holistic understanding of cancer. In the past decade, diverse integration methods have been developed, ranging from network-based\cite{wang2014similarity, rappoport2019nemo, nguyen2019pinsplus} and matrix factorization-based\cite{shen2012integrative, yang2016non} to Bayesian clustering techniques\cite{vaske2010inference, wu2015fast} and advanced deep learning approaches\cite{lee2021variational, chen2019deepmf}. Despite successes in disease subtyping\cite{de2018integration, stefanik2018brain} and advancing precision medicine\cite{hamamoto2019epigenetics},  these methods share a common limitation: the requirement for complete data across samples. This prerequisite becomes problematic due to the frequent occurrence of incomplete data in profiling assays, often a consequence of experimental or financial constraints. For instance, in integrating various genomic data types, it is common to have complete genotype information for all individuals, but gene expression and/or methylation data are frequently incomplete \cite{martin2017genomic}. Analyzing such incomplete omics data is challenging. Excluding samples with missing omics data significantly reduces sample sizes, particularly when integrating multiple omics layers, and imputing missing values can introduce bias and uncertainty\cite{little2019statistical, henry2013comparative}. This underscores the critical need for computational techniques capable of directly modeling heterogeneous multi-omics datasets “as is”, without requiring complete measurements or discarding useful information.


Advanced integrative methods to address the missing data issue can be classified into two categories:  joint imputation or optimization masking approaches \cite{flores2023missing}. Joint imputation approaches \cite{fang2018bayesian, argelaguet2020mofa+, lock2022bidimensional} predict missing values within the modeling framework, but the accuracy of these imputed values, which may introduce bias, is crucial to the results. Also, these approaches often require larger sample sizes for effective model estimation. On the other hand, optimization masking techniques\cite{lee2021variational, rappoport2019nemo, xu2021network, rappoport2020monet}, which work with processed data such as patient graphs, allow partial samples to contribute by masking missing data during the optimization process. Such approaches also have their own limitations. Some require the presence of at least one common data view across partially observed samples, which may not always be feasible\cite{rappoport2019nemo}.  Others grapple with increased computational complexity and potential inaccuracies in clustering outcomes as the number or size of graphs increases \cite{xu2021network, rappoport2020monet}.


Molecular subtypes identified through multi-omics integration offer essential diagnostic and prognostic insights. However, a major challenge in transitioning these integrative models to clinical practice lies in accurately classifying new patients into these predefined subtypes, particularly when dealing with incomplete omics data from these individuals\cite{hornung2023prediction}. This limitation significantly hinders the practical application of molecular subtypes in clinical settings, as patients often present with partial datasets that are not sufficiently addressed by current methodologies. The lack of robust computational approaches capable of making reliable predictions from these incomplete and diverse omics profiles is a critical barrier to the real-world clinical adoption of integrative models. Addressing this gap by developing methods that can infer accurate subtypes from any available data is essential for advancing personalized patient care and fully realizing the potential of multi-omics integration in medicine.

To overcome these limitations, we present \textbf{IntegrAO} (\textbf{Integr}ate \textbf{A}ny \textbf{O}mics), an unsupervised framework for integrating incomplete multi-omics profiles and classifying new samples with incomplete data. IntegrAO starts by integrating partially overlapped patient graphs derived from diverse omics data. Its unique partial graph fusion mechanism effectively enhances information integration with a high number of shared patients across modalities and adeptly adapts to situations with fewer overlapping samples. This capability allows IntegrAO to effectively combine diverse incomplete omics data, ensuring high fidelity and noise resistance. The framework then employs graph neural networks (GNNs) to extract and align patient embeddings from diverse raw omics features into a unified space. This unified embedding space is crucial for accurately classifying new patients into predefined subtypes using any available data, facilitating the transition to clinical practice. To demonstrate the use of IntegrAO, we first show IntegrAO exhibits robust integration of partially overlapping data across diverse missing data scenarios through simulation of omics dataset.  A case study in acute myeloid leukemia then illustrates IntegrAO's capacity to build a comprehensive view of heterogeneity from incomplete multi-omics.  Systematic evaluations conducted on five cancer cohorts, covering six omics modalities, underscore IntegrAO’s resilience to missing data and its effectiveness in integrating partial data and classifying new samples. Through its proficient handling of heterogeneous and incomplete datasets, IntegrAO emerges as a significant tool in precision oncology, facilitating an all-encompassing approach to patient characterization.

\section{Results}\label{sec2}

\subsection{IntegrAO Overview}\label{sec2}

\begin{figure}[!tp]
    \centering
    \includegraphics[width=12cm]{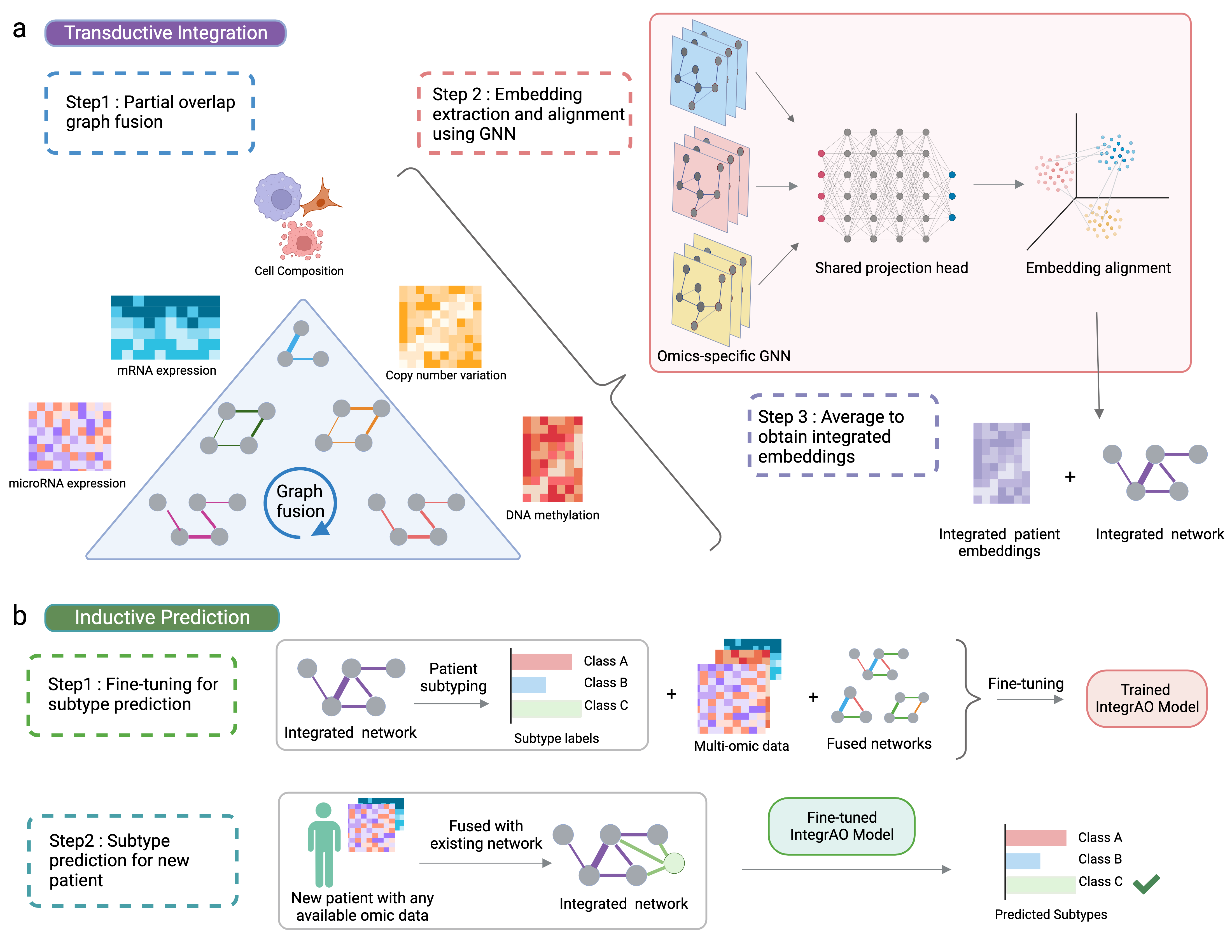}
    \caption{Overview of the IntegrAO framework. (a) Step 1: Example representation of cell composition, mRNA expression, microRNA expression, DNA methylation and copy number variation datasets are used to construct per-omics 
    patient graphs. Patient data need not encompass all omics types. Subsequently, a fusion phase iteratively refines each graph with information gathered from other graphs, culminating in a unified graph for each type of omics. Step 2: Both these unified graphs and their corresponding omics features are input into omics-specific Graph Neural Networks (GNNs) to learn patient embeddings. These low-dimensional patient embeddings are optimized to retain similarity information from the individual unified graphs while minimizing differences in embeddings for the same patients across different omics. Step 3: The conclusive embeddings are procured by averaging omics-specific embeddings and applied in the construction of the final integrated patient graph.  (b) Conversion of IntegrAO into a predictive framework. Utilizing the integrated graph, patient subtypes can be identified and leveraged to fine-tune the trained IntegrAO model. The fine-tuned IntegrAO model enables the classification of new patients with any accessible omics data. During the inference process, graph fusion is first conducted on new patients along with existing patients. The consequent fused graph and associated omics features are then input into the fine-tuned IntegrAO model, allowing for the prediction of patient subtypes.}
    \label{schema}
\end{figure}

We present IntegrAO, an unsupervised framework for integrating multi-omics datasets with partial overlap. As outlined in \textbf{Fig. \ref{schema}a}, IntegrAO has two key functionalities: transductive integration and inductive prediction. 

Transductive integration is structured around two core steps: (1) Fusion of partially overlapping patient graphs, (2) Unsupervised extraction and alignment of patient embeddings across omics modalities. In Step 1, IntegrAO is tailored to accommodate samples with missing data types. It first constructs a patient graph for each omic, with patients as nodes and weighted edges denoting pairwise similarities (Online Methods \ref{network construction}). IntegrAO then fuses graphs through an iterative update process utilizing a nonlinear method rooted in message-passing theory (Online Methods \ref{network fusion}). Notably, IntegrAO enables partial graph fusion by leveraging shared samples between omics - more shared patients increase information fusion, while fewer dampen it. By using common samples as bridges, patients with partial omics also get updated, enriching the composite profiles. As the extent of patient overlap may vary across each pair of omics data modalities, IntegrAO performs pairwise fusion between graphs to maximize the information flow. Step 1 yields a fused graph for each omic, encapsulating integrated information from other omics. Step 2 extracts low-dimensional patient embeddings from each omic into a unified space (Online Methods \ref{Embedding extraction}). The fused networks and omics data are fed into omic-specific graph neural network (GNN) encoders, then into a shared projection head to obtain embeddings per omic. The model training phase is designed to have the low-dimensional embeddings retraining similarity structures as the input fused graphs while simultaneously ensuring that embeddings of the same patients are aligned across different omics datasets. Final embeddings are obtained by averaging across omics to construct the integrated graph.

A key capability of IntegrAO is flexible transformation from unsupervised integration to supervised prediction (\textbf{Fig. \ref{schema}b}). Taking cancer patient subtyping as an example, once subtypes are discerned from the integrated graph, IntegrAO can be further fine-tuned to enable subtype prediction for new patients using any available omics data (Online Methods \ref{fine-tune}). The distinction between the prediction model and the unsupervised-training model lies in the added Multi-layer Perceptron(MLP) prediction head, which enables the processing of the averaged patient embeddings across omics for subtype prediction. While the prediction model inherits its initial weights from the unsupervised-training model, the MLP prediction head starts with a random weight initialization. Fine-tuning balances two key objectives: preserving the unsupervised objectives of patient embedding generation, and minimizing subtype classification loss. This dual optimization enables the model to support subtype prediction in a modality-agnostic manner. During the inference process, given any combination of multi-omic data for new patients, the first step involves fusing these new patients into existing graphs. Following this fusion, the fine-tuned model accepts the fused graphs along with the corresponding omics features, allowing IntegrAO to predict specific cancer subtypes (Online Methods \ref{model inference}). 


\subsection{Simulation: IntegrAO exhibits robust integration of partially overlapping data across diverse missing data scenarios}\label{sec2}

\begin{figure}[!tp]
    \centering
    \includegraphics[width=12cm]{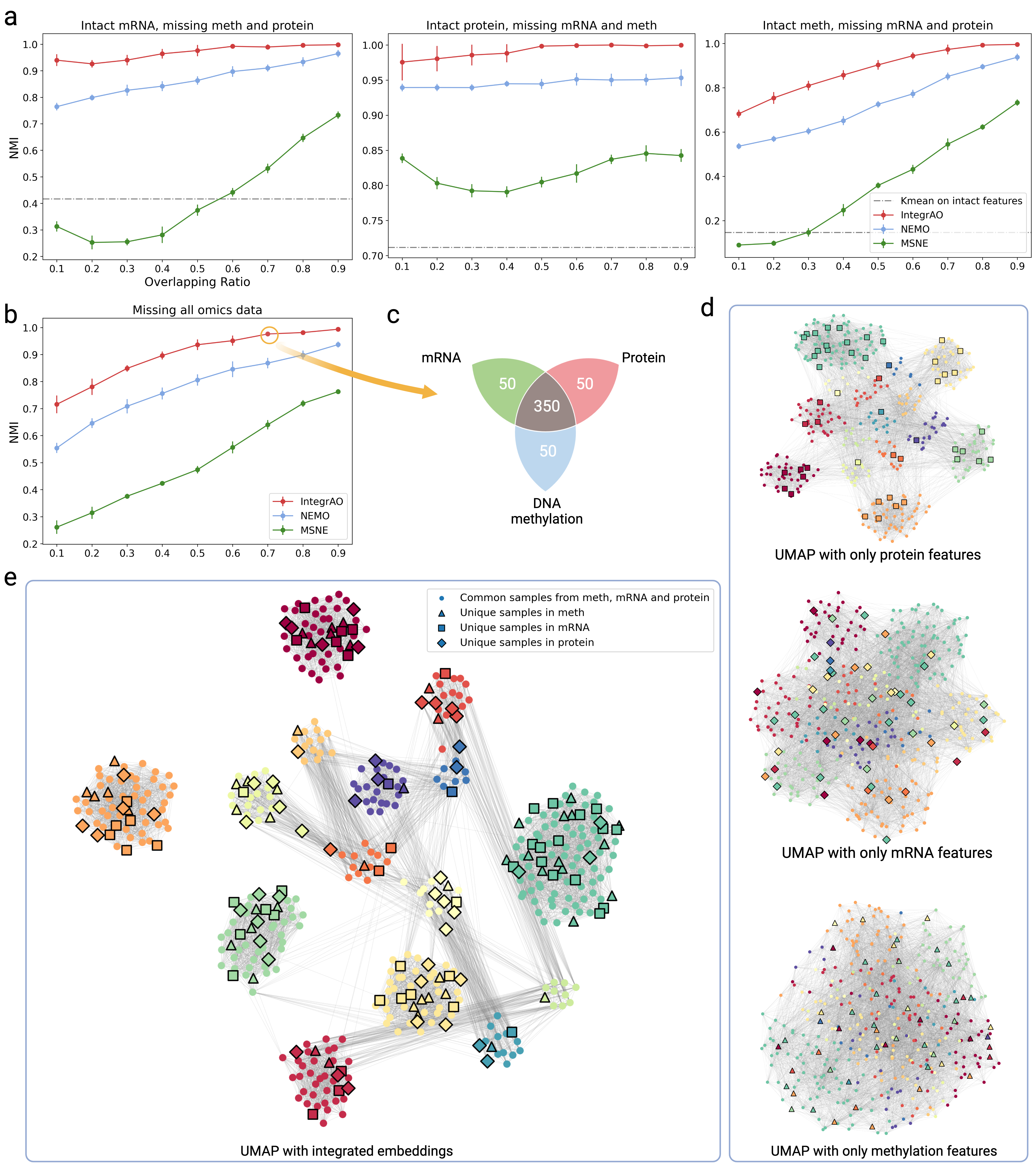}
    \caption{Benchmarking partial multi-omics integration between IntegrAO, NEMO, and MSNE on simulated multi-omics cancer dataset using Normalized Mutual Information (NMI). (a) NMI versus overlapping data ratio across three missing scenarios (n=10 experiments for each ratio). Means of evaluation metrics with standard deviations from different experiments are shown in the figure, where the error bar represents plus/minus one standard deviation. From left to right: Uniform random subsampling of DNA methylation and protein expression with intact mRNA expression; Uniform random subsampling of mRNA expression and DNA methylation with intact protein expression; Uniform random subsampling of mRNA expression and protein expression with intact DNA methylation. IntegrAO demonstrates superior performance in all scenarios. (b) IntegrAO outperforms other methods in a more challenging scenario where all omic data are partially missing. (c) An illustrative example with a 70\% data overlap ratio, showing 350 common and 50 unique samples per modality.  (d) Pre-integration UMAP visualizations for each modality for the 70\% all-missing data scenario, highlighting both common and unique samples. (e) Post-integration UMAP visualization of patient embeddings via IntegrAO. Upon integration, clustering resolution was enhanced with unique samples from each network showing improved alignment.}
    \label{simulate_omics}
\end{figure}

We first evaluated IntegrAO using a simulated multi-omics dataset generated by the \textit{InterSim} CRAN package \cite{chalise2016intersim}, which produces data for three omics modalities (DNA methylation, mRNA expression, and protein expression). We simulated a total of 500 samples with 15 clusters, and each cluster have variable random sizes (Online Methods \ref{data processing}).  IntegrAO was compared to two related network-based methods capable of handling partial overlap, NEighborhood-based Multi-Omics clustering (NEMO)\cite{rappoport2019nemo} and Multiple Similarity Network Embedding (MSNE)\cite{xu2021network}, using Normalized Mutual Information (NMI) to assess clustering congruence with ground truth labels. NEMO and MSNE were run using their default settings and hyperparameters. We first tested the scenarios where one omic modality remains intact and two other modalities undergo uniform random subsampling at ratios from 0.1 to 0.9 (\textbf{Fig. \ref{simulate_omics}a}). The random sub-sampling process was repeated 10 times for each overlapping ratio. 

In integration scenarios with partial overlap, two regimes emerge: low overlap, where the goal is to minimize inter-modality influence due to potential noise from limited shared samples, and high overlap, where the objective is to maximize information flow between modalities as the increased common samples enable more reliable integration between modalities. IntegrAO substantially outperformed other methods across all overlap ratios and maintained robust performance even in low-overlap situations where MSNE faltered (\textbf{Fig. \ref{simulate_omics}a}). In the latter scenario, MSNE's performance can decline when integrating more data, sometimes falling below baseline levels established by K-means clustering on the intact modality, highlighting its limitations in handling low data overlap with noise signals from other modalities. We subsequently evaluated a more complex experimental setup in which no omic data modality remained intact. In this experiment, we first selected a subset of common samples based on the specified overlapping ratio, then evenly distributed the remaining samples among the three modalities as unique entities. We observed enhanced clustering performance in all three methods as the overlapping ratios increased. IntegrAO consistently outperformed other methods, maintaining effective clustering even at a minimal 10\% overlap (\textbf{Fig. \ref{simulate_omics}b}). The superior performance of IntegrAO can be attributed to its ability to fuse unique samples even just with their individual modality, in contrast to NEMO, which requires samples to be observed in at least one common view with others. This distinction highlights IntegrAO's proficiency in utilizing unique and incomplete datasets, effectively extracting valuable information where other methods may fall short.

To further investigate IntegrAO's integration effectiveness, we conducted a detailed visual analysis on a 70\% overlap scenario with 350 shared samples and 50 unique samples per modality (\textbf{Fig. \ref{simulate_omics}c}). We generated UMAP visualizations for each omic type prior to integration. In these visualizations, dots represented shared samples, while diamonds, squares, and triangles indicated unique samples of mRNA, protein, and DNA methylation, respectively (\textbf{Fig. \ref{simulate_omics}d}). Pre-integration, the embeddings displayed an entangled structure with randomly dispersed unique samples. In contrast, following IntegrAO integration, the UMAP shows clearly defined clustering of the 15 clusters, with coherent grouping of unique samples (\textbf{Fig. \ref{simulate_omics}e}). This highlights IntegrAO's ability to disentangle complex mixed signals and uncover integrated structures through joint analysis of distinct but partially overlapping datasets.


\subsection{IntegrAO identifies fine-grained clinically and biologically distinct AML subtypes}\label{sec2}

\begin{figure}[!tp]
    \centering
    \includegraphics[width=12cm]{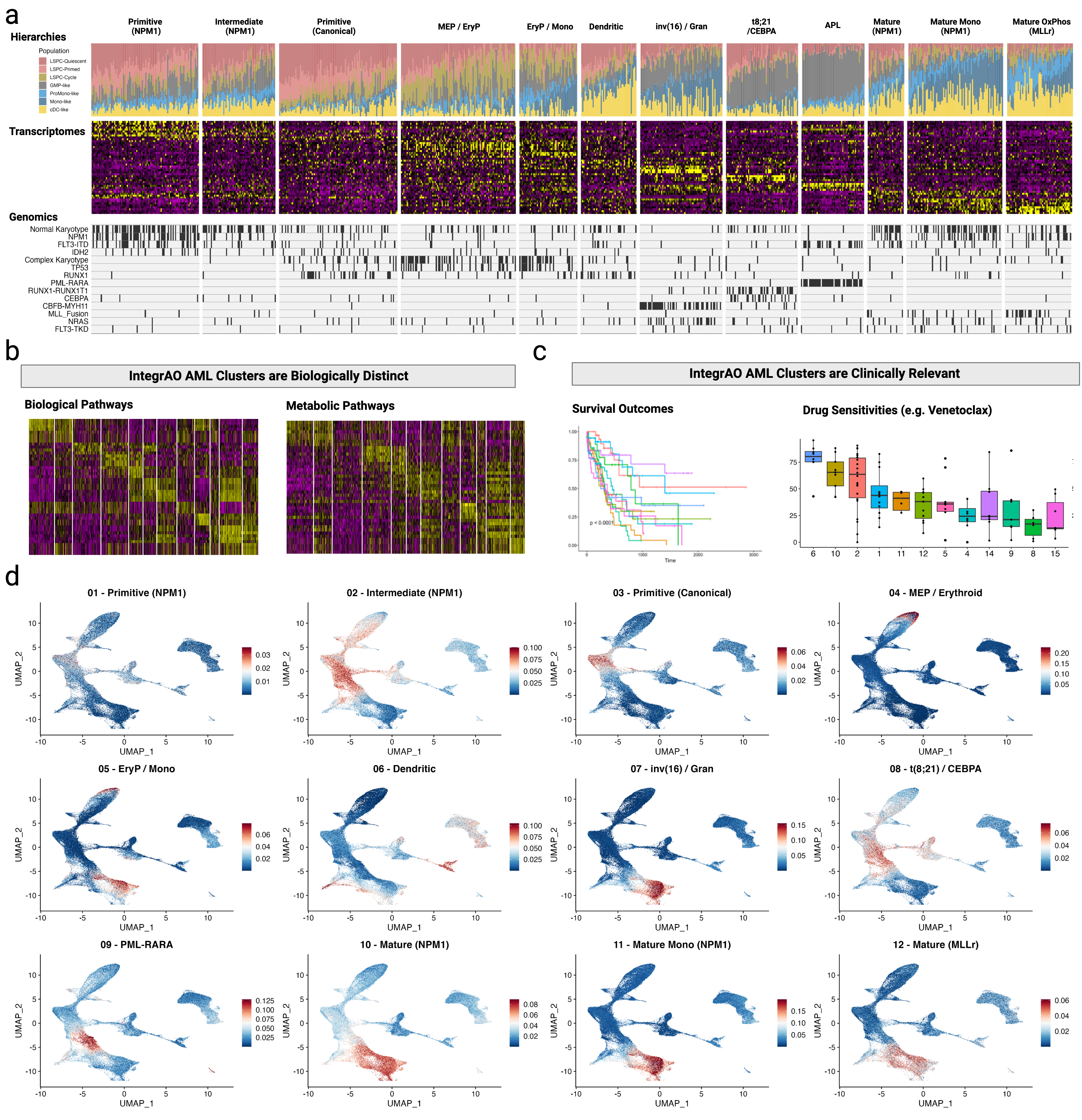}
    \caption{ Multi-omics integrative analysis of acute myeloid leukemia (AML) elucidating intertumor heterogeneity. (a) IntegrAO discerns 12 subtypes with distinct hierarchical composition, transcriptomic profiles, and mutational patterns, preserving granular differentiations. (b) IntegrAO subtypes demonstrate greater differential survival versus individual datasets. (c) More significantly sensitive drugs are revealed by IntegrAO versus single data types. (d) Hematopoietic lineage enrichment analysis validates subtype differentiation, underscoring captured heterogeneity. }
    \label{aml_result}
\end{figure}

To elucidate heterogeneity in acute myeloid leukemia (AML), a cancer marked by extensive inter-patient and intra-patient heterogeneity, we applied IntegrAO to an empirical AML dataset. Recently, a new layer of heterogeneity has been identified in AML corresponding to the composition of each patient’s leukemia cell hierarchy\cite{zeng2022cellular}, providing new insights into disease biology and drug response. We sought to utilize IntegrAO to integrate this new information with two other modalities, mRNA expression and DNA methylation, to achieve an unprecedented multi-dimensional perspective on AML heterogeneity. We thus applied IntegrAO to three AML cohorts, TCGA, BEAT-AML\cite{tyner2018functional}, and Leucegene\cite{marquis2018high}, leveraging mRNA expression and hierarchy composition for 812 patients, and methylation profiles from 308 patients of those patients (Online Methods \ref{data processing}).

IntegrAO integration of mRNA, DNA methylation, and cell hierarchy data revealed 12 biologically distinct AML subtypes (Online Methods \ref{choose number}), exhibiting unique multi-omics patterns that provide a refined resolution of heterogeneity, as shown in \textbf{Fig. \ref{aml_result}a and Supplementary Fig. S1}. Notably, the subtypes refine broader groupings defined previously using only the hierarchy data by \textit{Zeng et al}\cite{zeng2022cellular}, validating IntegrAO's capacity to extract nuanced diversity. Detailed examination of cell compositions supports this, with `Primitive' subtypes enriched for primitive leukemia stem and progenitor cells (LSPCs), `t8;21/CEBPA' and `APL' enriched for GMP-like cells, and `Mature' subtypes for Mono-like and cDC-like cells. Despite similar compositions, the two `Primitive' subtypes are differentiated by distinct mutations - `Primitive (NPM1)' associated with NPM1/FLT3-ITD alterations, `Primitive (Canonical)' with TP53/RUNX1. Further heterogeneity is observed in the four NPM1-driven subtypes with divergent hierarchies. Notably, a novel subtype emerged dominated by erythroid progenitor (EryP) cells, a finding that diverges from conventional understanding and may inform future AML research directions. These granular insights highlight IntegrAO's effectiveness in eliciting nuanced biology underlying AML diversity, potentially informing tailored therapeutic strategies. Furthermore, the heatmaps derived from Gene Ontology (GO) analysis of biological and metabolic pathways closely mirror the subtypes identified by IntegrAO, highlighting their biological significance and uniqueness (\textbf{Fig. \ref{aml_result}b}). In the GO biological pathways heatmap, distinct segments correspond to specific subtypes, reflecting dominant biological processes such as cellular functions, regulatory mechanisms, and interaction pathways. Likewise, the GO metabolic pathways heatmap clearly segments into areas representing key metabolic activities, including glycolysis, lipid metabolism, and energy production, characteristic of these clusters. Additionally, we conducted VIPER analyses on both all regulons and transcription factor-specific regulons, with the resulting heatmaps demonstrating distinct block structures that align well with the clusters defined by IntegrAO (\textbf{Supplementary Figs. S2-3}). Collectively, these results further emphasize the biological distinctness of the subtypes identified by IntegrAO.

We further assessed the clinical importance of the subtypes through survival analysis and drug sensitivity profiling. Kaplan-Meier survival curves for the clusters, drawn from the combined TCGA and BEAT-AML cohort, showed significant differences (multi-group logrank test p-value = 1.21e-7) (\textbf{Fig. \ref{aml_result}c}). Separate analysis of TCGA and BEAT-AML data also revealed significant survival distinctions (\textbf{Supplementary Figs. S4a,b}). We also conducted nested likelihood ratio tests to determine whether the addition of subtype clustering enhances prognostic stratification beyond four established factors (age, cytogenetic risk, white blood cell count, and NPM1 mutation). Compared to subtypes identified using only mRNA or cell hierarchy data, IntegrAO subtypes showed greater multivariate prognostic significance (p-value=0.01425), while subtypes from individual data types did not demonstrate significant improvement (p>0.05) (\textbf{Supplementary Fig. S4c}). For drug sensitivity, ANOVA tests were used to assess whether IntegrAO subtypes show differential responses to each of 122 anti-cancer agents in the BEAT AML drug screening dataset (\textbf{Fig. \ref{aml_result}c}). A differential response was indicated by an ANOVA p-value < 0.05. The analysis revealed that 47 out of the 122 drugs showed differential sensitivity in IntegrAO clusters, affirming their clinical utility (\textbf{Supplementary Fig. S5}).

To validate the heterogeneity captured in the IntegrAO AML subtypes, we evaluated the enrichment along defined stages of hematopoietic differentiation in each defined subtype (\textbf{Fig \ref{aml_result}d}). As a reference, we utilized the single-cell UMAP of bone marrow mononuclear cells from \textit{Galen et al}\cite{van2019single}, providing an unbiased landscape of normal hematopoietic differentiation (\textbf{Supplementary Fig. S6}). We then mapped the specific populations most enriched in each IntegrAO subtype onto this independent reference. Notably, this revealed alignments including the `Dendritic' subtype with plasmacytoid and conventional dendritic cells, `Primitive (Canonical)' with hematopoietic stem cells, and `Mature Mono (NPM1)' with monocytes, etc. The orthogonal validation that IntegrAO subtypes align with varying normal developmental trajectories highlights that IntegrAO integration preserves, and does not smooth over, the heterogeneous lineages underlying AML intertumor heterogeneity. 

In summary, IntegrAO integration of complete and incomplete AML data effectively identifies distinct subtypes with biological and clinical relevance. By effectively sharing information yet preserving essential distinctions between omics, IntegrAO offers a comprehensive insight into cancer complexity, furthering biological discovery and precision medicine. IntegrAO's development of such detailed patient stratification that correlates with clinical outcomes and biological underpinnings demonstrates its potential in guiding individualized therapeutic decisions, especially in complex conditions like AML.


\subsection{A Pan-Cancer Evaluation of IntegrAO on identifying clinically distinct subtypes}\label{sec2}

To further evaluate the efficacy of partial multi-omics integration, a comparative analysis was conducted between IntegrAO, NEMO, and MSEN across five distinct cancer datasets sourced from The Cancer Genome Atlas (TCGA)\cite{cancer2008comprehensive}. For each cancer type, we leveraged the maximum number of patients in each of the five omics: mRNA expression, DNA methylation, miRNA expression, Reverse-phase protein array, and copy-number variation. By utilizing all possible samples from TCGA, this benchmark dataset encompasses rich, heterogeneous profiles without data waste. Recently, cell composition derived from deconvolving bulk mRNA expression data has emerged as a critical modality for the delineation of disease subtypes and the tailoring of therapeutic strategies. Uniquely, we additionally incorporated cell type composition, as an extra modality to enhance heterogeneity characterization (Online Methods \ref{cell deconvolution}). The details of data collection and preprocessing can be found in Online Methods \ref{data processing}. The respective patient counts and feature counts for each modality are detailed in \textbf{Supplementary Tables S1-2}. As the integration of additional modalities progresses, acquiring a sufficient number of common samples across all views becomes increasingly challenging. Consequently, the ability to integrate partial omics datasets is essential, allowing for the efficient utilization of all existing data without squandering valuable information. 

\begin{figure}[!tp]
    \centering
    \includegraphics[width=12cm]{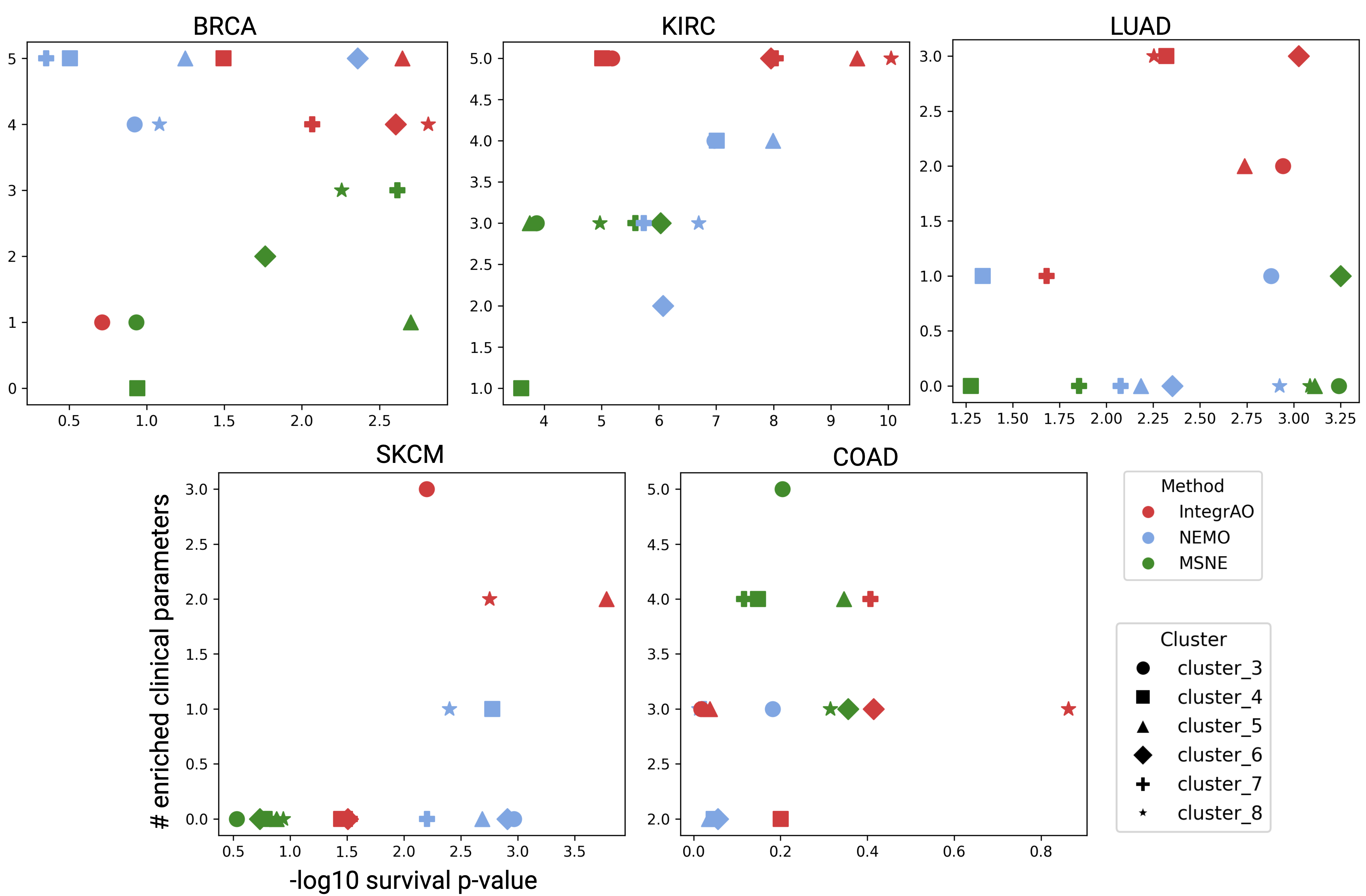}
    \caption{Comparative analysis of IntegrAO, NEMO, and MSNE across 5 cancer types with partial multi-omics data. The x-axis depicts differential survival between clusters, quantified by -log10 of the P-value from age-adjusted nested log-rank testing (higher indicates greater survival differentiation). The y-axis shows the number of enriched clinical parameters within clusters (higher denotes more parameters enriched). Each plot compares methods for a cancer dataset for different cluster numbers. Overall, IntegrAO more reliably identifies clusters with both better survival differentiation and higher clinical enrichment than other methods.}
    \label{cancer result}
\end{figure}

To evaluate the effectiveness of a given clustering solution, two specific metrics were employed. First, age-adjusted differential survival between the resultant clusters was measured using the logrank test. This method operates on the premise that clusters with significant differences in survival rates reflect biologically meaningful variations. Subsequently, we examined the enrichment of six clinical labels within the clusters, including gender, age at diagnosis, pathologic T (tumor progression), pathologic M (metastases), pathologic N (cancer in lymph nodes), and pathologic stage (total progression). Enrichment for discrete parameters was assessed using the $\chi^2$ test for independence, while numeric parameters were evaluated using the Kruskal-Wallis test. Recognizing the absence of a definitive ground truth for the number of clusters pertaining to each cancer type, we executed clustering for a range of cluster numbers from 3 to 8. \textbf{Fig. \ref{cancer result}} illustrates the comparative performance of IntegrAO against other methods across various cancer datasets.

Overall, IntegrAO reliably identified subtypes with both superior survival differentiation and clinical variable enrichment across cancer cohorts. In BRCA, KIRC, and SKCM, IntegrAO solutions were clearly favorable considering both criteria. In LUAD, IntegrAO achieved significantly better clinical enrichment despite comparable survival differentiation to MSNE. And for COAD, IntegrAO showed better survival stratification amongst methods despite suboptimal clinical enrichment results. In contrast, NEMO and MSNE demonstrated inconsistent performance across cancer types. MSNE delivered satisfactory results in COAD, yet its performance was less convincing in KIRC and SKCM. Meanwhile, NEMO showcased a strong performance in BRCA, but this did not extend to COAD or LUAD. Furthermore, the uneven ability of MSNE in discerning survival differences—evident in BRCA but absent in KIRC and SKCM—alongside NEMO's variable success in pinpointing clinically enriched variables, with success in BRCA but not in COAD or LUAD, highlights a significant shortfall. IntegrAO proficiently discerned both criteria, reflecting robust integration and patient stratification. This inconsistency among the other methods underscores the intricate challenge of integrating diverse partial multi-omics data, which also underscores IntegrAO’s importance for translational applications requiring holistic patient characterization.


\subsection{IntegrAO enables robust new patient classification using incomplete omic-data}\label{sec2}

\begin{figure}[!tp]
    \centering
    \includegraphics[width=12cm]{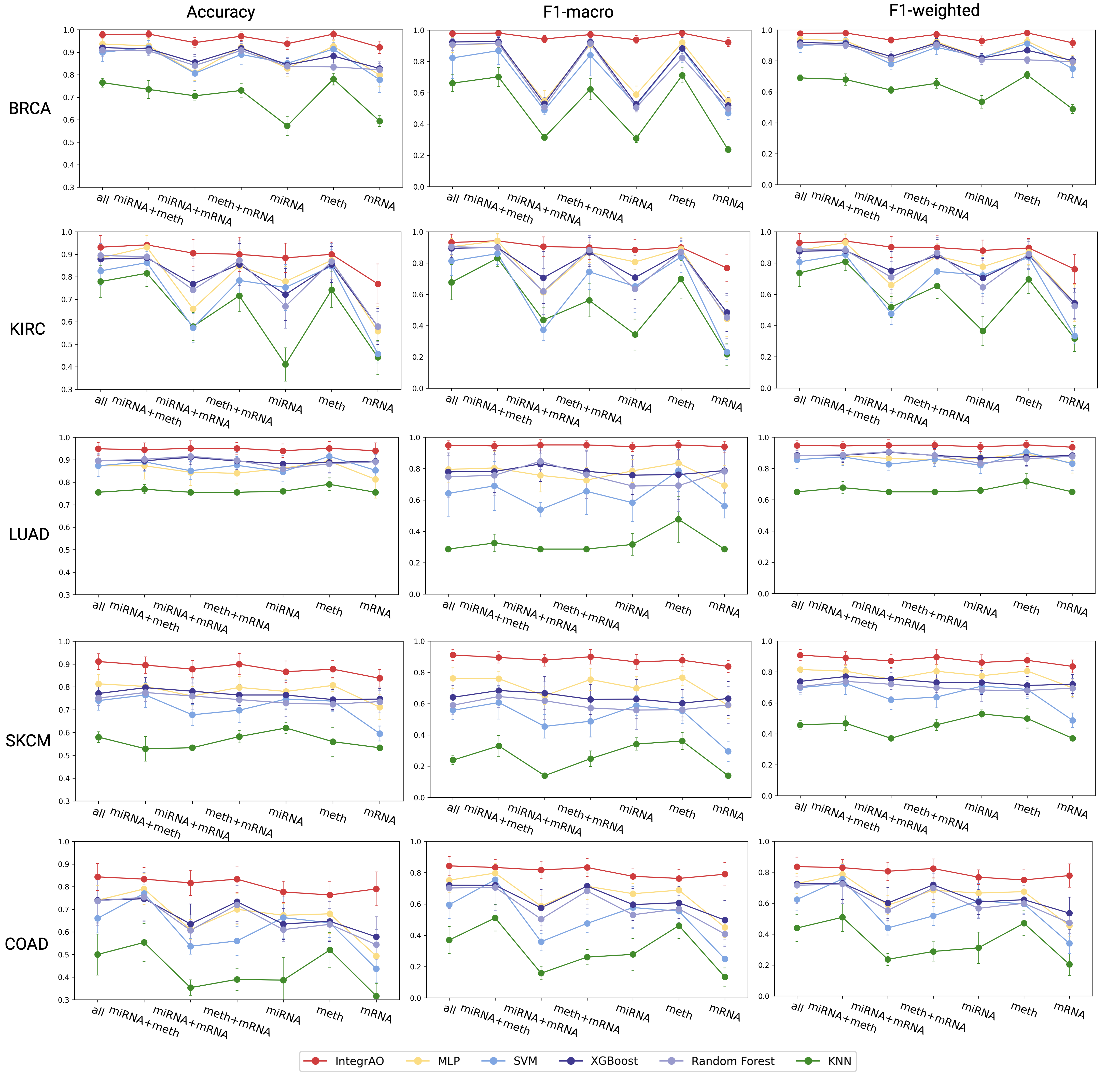}
    \caption{ Performance comparison of new patient classification using IntegrAO versus MLP, SVM, XGBoost, Random Forest, and KNN under different omic combinations. Accuracy, F1-macro, and F1-weighted were evaluated, with means and standard deviations from multiple experiments displayed (error bars denote ±1 standard deviation). mRNA, meth, and miRNA refer to single-omic classification using mRNA expression, DNA methylation, and miRNA expression data respectively. miRNA+meth, miRNA+mRNA, and meth+mRNA indicate classification with two omics, while "all" used all three data types. Across all metrics and inputs, IntegrAO substantially outperforms other methods, highlighting its ability to effectively leverage diverse omics for integrative patient classification. }
    \label{cancer_projection}
\end{figure}

In clinical applications, after discerning patient subtypes, categorizing new patients into predefined clusters is often needed but overlooked by many methods. This task is more complex when new patients possess only partial omics data. Thus, methodologies that can classify new samples lacking comprehensive features are critical. IntegrAO enables new patient classification into established subtypes using any available omics data. This key functionality addresses an important unmet need for translating integrative methods into precision medicine applications.

To rigorously assess IntegrAO's proficiency in classifying new patients, we designed an experimental framework that mimics the real-world scenario of assigning unseen samples to predefined subtypes. We benchmarked IntegrAO against five widely-used classifiers: Multi-layer Perceptron (MLP), Support Vector Machine (SVM), Random Forest, XGBoost, and K-Nearest Neighbors (KNN). Our ground truth dataset was derived from comprehensive multi-omics data, including miRNA, mRNA, and DNA methylation profiles from the five TCGA cancer cohorts, selecting only patients with a full set of data across these modalities. IntegrAO was employed to integrate the complete dataset to construct an integrated network, which was then used to determine the optimal number of clusters and generate cluster labels via spectral clustering (Online Methods \ref{choose number} and \textbf{Supplementary Fig. S7}). The optimal number of clusters for each cancer type is listed in \textbf{Supplementary Table S3}. This full cohort was then utilized in a rigorous stratified 10-fold cross-validation procedure. In each fold, the methods were trained on 90\% of samples and subsequently tested on the held-out 10\% of unseen samples. To assess multi-class prediction performance, accuracy, F1-macro, and F1-weighted were measured as key evaluation metrics. For each dataset, IntegrAO first conducted unsupervised integration on the 90\% training samples to discern subtypes, then refined the model using the known "ground truth" labels, and finally employed the fine-tuned model to predict the subtype of the unseen test samples using any combination of the omics data. In contrast, the other methods were trained on either single omics or direct concatenated multi-omics data from the 90\% subset, and evaluated on their ability to correctly predict the subtypes of the 10\% held-out test set.

IntegrAO consistently and substantially outperformed all comparative classification methods across every new patient projection task, as quantified by accuracy, F1-macro, and F1-weighted metrics (\textbf{Fig. \ref{cancer_projection}}). In particular, IntegrAO demonstrated clearly superior performance, while KNN was notably the least effective, and the remaining algorithms exhibited intermediate but significantly inferior accuracy compared to IntegrAO. Further analysis revealed that IntegrAO's classification performance was highly robust across diverse omic combinations, whereas other methods displayed pronounced fluctuation and instability when missing certain data modalities. This instability arises because specific integrated omics can be highly noisy or misleading for overall subtyping. Classifying new patients with only that noisy modality is then extremely challenging to accurately map into the defined subtypes. IntegrAO overcomes this by embedding different omic features into a unified space, enabling it to approximate the classification accuracy of full multi-omics datasets even with incomplete data. This feature holds significant clinical importance, as physicians frequently face the challenge of making diagnostic or treatment decisions with only partial omic information available. By effectively bridging this gap, IntegrAO emerges as a pivotal tool that enhances the application of multi-omics approaches in the practical landscape of precision medicine, facilitating better-informed clinical decisions.


\section{Discussion}\label{discussion}

This study presents IntegrAO, an integrative framework designed to tackle key challenges in multi-omics analysis - handling incomplete heterogeneous data and projecting new samples using partial profiles. The results validate IntegrAO's ability to integrate diverse cancer datasets with missing modalities and to classify new patients reliably. Tests with simulated cancer omics data reveal IntegrAO’s capability to integrate missing data in various scenarios, showing resilience to noise at low data overlaps and effective integration at higher overlaps. In the case study on acute myeloid leukemia, IntegrAO successfully combined cell hierarchy composition, transcriptomics, and DNA methylation, identifying 12 clinically and biologically distinct subtypes and illustrating AML’s heterogeneity. Systematic evaluations across five cancer cohorts, encompassing six omics modalities, show IntegrAO’s superiority in identifying significant subtypes compared to other methods. Its consistent performance in projecting new samples, regardless of the number of available omics, highlights its potential in modality-agnostic inference and unified patient representation.

IntegrAO stands out in its ability to handle varied and incomplete data sets, establishing itself as a pivotal tool for the future of precision medicine.  Its architecture is specifically tailored to not only accommodate but also to synergize disparate data types, thereby maximizing the utility of every available data point. This aspect is particularly crucial in clinical settings, where data availability can often be unpredictable and inconsistent. IntegrAO's ability to integrate these disparate data into a unified space represents a significant advancement in patient care. Furthermore, IntegrAO's ability to predict outcomes from new and incomplete samples lays the groundwork for the practical application of integrative models in clinical settings, including diagnosis and personalized treatment. Through these capabilities, IntegrAO is revolutionizing the creation and use of comprehensive patient databases. It enables a more nuanced understanding of cancer and facilitates a seamless transition of these insights into clinical practice.

This research lays the foundation for several crucial future developments to enhance IntegrAO into a robust, scalable, and broadly applicable integrative framework. A key step is transforming the graph fusion process into an end-to-end deep neural network, critical for enhancing scalability and flexibility when analyzing massive biomedical datasets. Additionally, incorporating diverse data types, such as histopathology images, clinical notes, and sensor data, will allow for more detailed profiling and subtyping. Moreover, potential areas of application extend beyond cancer patient stratification to cell subtyping, drug discovery, biomarker identification, and precision nutrition. Conducting thorough evaluations across various applications, alongside efforts to enhance model interpretability, is crucial for showcasing IntegrAO's utility and reliability in various biomedical domains. By pushing boundaries on multiple fronts, this work paves the way for positioning IntegrAO as a crucial model for the future of precision medicine.


\section{Methods}\label{sec6}

\subsection{Data Preprocessing}\label{data processing}

\paragraph{Simulated cancer omics datasets}
We utilized the \textit{InterSim} CRAN package\cite{chalise2016intersim} to simulate cancer omics datasets, generating a total of 500 samples distributed across 15 clusters of varying sizes, reflecting realistic clinical scenarios. For the hyperparameters, we set 'effect=0.1' and 'p.DMP=0.1', while keeping the rest of the hyperparameters at their default values.

\paragraph{TCGA cancer datasets} For the cancer datasets, we leveraged multi-omic data across five tumor types from The Cancer Genome Atlas (TCGA) - breast invasive carcinoma (BRCA), colon adenocarcinoma (COAD), skin cutaneous melanoma (SKCM), kidney renal clear cell carcinoma (KIRC), and lung adenocarcinoma (LUAD). Specifically, we obtained mRNA expression, DNA methylation, copy number variation, and protein expression data directly from cBioportal. MicroRNA expression data was retrieved separately from the Broad Institute's Firehose source data. Relevant clinical information was also acquired for each patient. Before analysis, rigorous preprocessing was performed, including outlier removal, imputation of missing values via k-nearest neighbors (kNN), and normalization by standard scaling to mean 0 and standard deviation 1. Patients with over 20\% missing data for any data type and features with over 20\% missing values across patients were excluded. We additionally selected the top 2,000 features exhibiting the greatest standard deviation from each data modality. For modalities with fewer than 2,000 total features, no feature filtering was performed.

\paragraph{AML cancer dataset}
To construct an integrated AML dataset for heterogeneous analysis, we merged raw data from the TCGA, BEAT-AML, and Leucegene cohorts. Gene expression data normalization was performed using a variance-stabilizing transformation for each each dataset. Batch effects were then corrected with the One Cell at A Time (OCAT)\cite{wang2022one} algorithm, which also reduced the features to a 30-dimensional space. For cell composition, we employed bulk gene expression deconvolution following \textit{Zeng et al.}\cite{zeng2022cellular}, applying OCAT for subsequent feature reduction. DNA methylation data, exclusive to the TCGA cohort, required no batch correction, and we selected 2,000 highly variable features based on dispersion. The final dataset included 812 AML patients with cell hierarchy composition and mRNA expression data, and a subset of 308 patients with additional DNA methylation data.

\subsection{Transductive Integration - Graph fusion}\label{network construction}

The first step of IntegrAO's transductive integration is the fusion of partially overlapping patient graphs. The subsequent section details the construction of these patient graphs and their partial overlap fusion. This graph fusion approach builds upon our prior work, Similarity Network Fusion (SNF)\cite{wang2014similarity}.

\paragraph{Patient graph construction}

We first construct a patient graph for each omic. Each graph can be represented as $G = (V, E)$, with vertices $V$ correspond to the patients $\{x_{1}, x_{2}, ..., x_{n}\}$ and undirected weighted edges $E$ denote the affinity between patients. The weight of the edge is computed with:

\begin{equation}\label{matrix_sim}
    W(i,j) =  \mbox{exp}\left(\frac{\rho^{2}(x_{i}, x_{j})}{\mu\varepsilon_{i, j}}\right),
\end{equation}
where $\rho(x_{i}, x_{j})$ is the Euclidean distance between patients $x_{i}$ and $x_{j}$. $\mu$ is a hyperparameter that is recommended setting in the range of [0.3, 0.8]. $\varepsilon_{i, j} $ is defined as 

\begin{equation}\label{euclidean_distance}
    \varepsilon_(i,j) = \frac{1}{3} \cdot \left( \frac{1}{\vert N_{i} \vert} \sum_{k \in N_{i}}\rho(x_{i}, x_{k}) + \frac{1}{\vert N_{j} \vert} \sum_{l \in N_{j}}\rho(x_{j}, x_{l})+ \rho(x_{i}, x_{j}) \right),
\end{equation}

where $N_{i}$ is the set of $x_{i}$'s neighbor including $x_{i}$ in $G$. We then performed two operations on each graph to derive the transition probability matrix for the graph fusion stage: the first is normalizing the affinity matrix for numerical stability:

\begin{equation}\label{stable_norm}
    P(i,j) =  \begin{cases}
    \frac{W(i, j)}{2\sum_{k \neq i} W(i, k)}, \hspace{0.3cm} i \neq j \\
    \hspace{1.4 cm} 1/2, \hspace{0.3cm} i = j \\
    \end{cases}.
\end{equation}
And the second is obtaining the local affinity matrix by considering only the K most similar patients per patient:

\begin{equation}\label{local_sim}
    S(i,j) =  \begin{cases}
    \frac{W(i, j)}{\sum_{k \in N_{i}} W(i, k)}, \hspace{0.3cm} j \in N_{i} \\
    \hspace{1.4 cm} 0, \hspace{0.3cm} \mbox{otherwise} \\
    \end{cases}.
\end{equation}
Given $v$ different data modalities, we can construct affinity matrices $W^{(m)}$ using Eq. \ref{matrix_sim} for the $m^{th}$ view, m = 1,2,..., v. $P^{(m)}$ and $S^{(m)}$ are obtained from Eq. \ref{stable_norm} and \ref{local_sim} respectively.

\paragraph{Partial overlap graph fusion}\label{network fusion}

In the case of two modalities with partially overlapping patient sets, i.e., $v = 2$, let $a$, $b$ denote the total number of patients for each modality, respectively, and $c$ the number of common patients. Let $C$ denote the set of common patients. The transition probability matrices $P^{(1)} \in \mathbb{R}^{a \times a} $ and $P^{(2)} \in \mathbb{R}^{b \times b}$, and local affinity matrices $S^{(1)} \in \mathbb{R}^{a \times a}$ and $S^{(2)} \in \mathbb{R}^{b \times b}$ are constructed as described previously. During fusion, each modality patient graph is initialized to its $P$ matrix ($P_{t=0}^{(1)} = P^{(1)}$; $P_{t=0}^{(2)} = P^{(2)}$). The key concept for fusing such partially overlapped data is to leverage the common samples to propagate information across the graphs via graph fusion. IntegrAO iteratively updates the patient graph for each data modality as follows:
\begin{equation}\label{fusion_1}
    P_{t+1}^{(1)} =  S^{(1)} \times P_{t}^{'(2\xrightarrow{}1)} \times (S^{(1)})^T,
\end{equation}
\begin{equation}\label{fusion_2}
    P_{t+1}^{(2)} = S^{(2)} \times P_{t}^{'(1\xrightarrow{}2)} \times (S^{(2)})^T,
\end{equation}
where the intermediate transition matrices $P_{t}^{'(2\xrightarrow{}1)}$ and $P_{t}^{'(1\xrightarrow{}2)}$ is obtained by first getting the affinity weights from the other modality of the common samples, as: 
\begin{equation}\label{partial_fusion_1}
    \underset{(a \times a)}{W_{t}^{'(2\xrightarrow{}1)}(i,j)} =  \begin{cases}
    P_{t}^{(2)}(i,j), \hspace{0.3cm} i, j \in C \\
    \hspace{1.2 cm} 0, \hspace{0.3cm} \mbox{otherwise} \\
    \end{cases},
\end{equation}
\begin{equation}\label{partial_fusion_2}
    \underset{(b \times b)}{W_{t}^{'(1\xrightarrow{}2)}(i,j)} =  \begin{cases}
    P_{t}^{(1)}(i,j), \hspace{0.3cm} i, j \in C \\
    \hspace{1.2 cm} 0, \hspace{0.3cm} \mbox{otherwise} \\
    \end{cases}.
\end{equation}
Then we apply a novel scaling normalization:
\begin{equation}\label{scale_norm}
\begin{split}
    P_{t}^{'}(i,j) =  \begin{cases}
    \frac{W_{t}^{'}(i,j)}{2\sum_{k \neq i} W_{t}^{'}(i,k)} \cdot \tau, \hspace{0.3cm} i \neq j \\
    \hspace{0.9 cm} 1 - 1/2 \cdot \tau, \hspace{0.3cm} i = j \\
    \end{cases}, \\
    \tau = \frac{\mbox{c}}{\mbox{number of sample in the current network}}.
\end{split}
\end{equation} During the iterative updates, each modality utilizes the shared patients' transition matrix from the other modality for fusion. The scaling normalization helps minimize the impact of the other modality when few patients are shared, while maximizing information flow when many patients are common. Not only the common patients' similarities can get updated through graph fusion, but the unique patients can also leverage the affinity information of the common patients from other modalities to learn more robust affinity for their own patient graph. This procedure updates the transition matrices each time generating two parallel interchanging fusion processes. After each iteration, we performed normalization on $P_{t+1}^{(1)}$  and $P_{t+1}^{(2)}$  as in Eq. \ref{stable_norm}, for the following three reasons: (i) ensure a patient is always most similar to themself than to other patients; (ii) ensure the final graph is full rank; (iii) for quicker convergence of fusion. After $t$ steps, we obtain the fused patient graph for each modality.

As our fusion approach leverages shared patients between modalities, the number of common patients may decrease when integrating more than two data types ($v >2 $). To address this, we perform pairwise fusion for multi-modalities following Eq. \ref{fusion_1} and \ref{fusion_2}:
\begin{equation}\label{multi_view_fusion}
    P^{(m)} =  \frac{\sum_{k\neq m}(S^{(m)} \times P^{'(k)} \times (S^{(m)})^T )}{v-1}, m = 1, 2, ..., v.
\end{equation}
Since the sample size differs across modalities, the fused affinity matrices for each data type retain the original dimensionality. The subsequent step involves integrating these modal-specific graphs into a unified representation, which will be detailed in the following section.

\subsection{Transductive Integration - Embedding extraction and alignment}\label{Embedding extraction}

The second step of IntegrAO's transductive integration is unsupervised extraction and alignment of patient embeddings across omics modalities. This embedding step fulfills two critical goals: (i) deriving low-dimensional embeddings that maintain the affinity structure of the fused graphs for each data type, and (ii) aligning embeddings for the same patient across modalities. 

\paragraph{Model architechture}

The deep learning model in IntegrAO consists of two key components: (1) Omic-specific graph encoders to extract patient embeddings within each data modality, (2) Shared projection layers to map the embeddings from different omics into a common latent space. For each omic-specific GNN encoder, inspired by GraphSAGE\cite{hamilton2017inductive}, instead of training individual embeddings for each node, we learn an aggregating function that generates embeddings by aggregating features from a node's local neighborhood. This enables generating embeddings for unseen nodes using the learned functions given their local neighborhood is provided. 

Using the fused patient graphs, we obtain sparse affinity matrices per omic by considering only the $K$ most connected neighbors of each patient node as defined in Eq. \ref{local_sim}. The weighted graphs are converted to unweighted versions as inputs to the encoders. Formally, let $G=(V, E)$ denote the unweighted patient networks, where $V$ are the nodes (patients) and $E$ are the edge connections (patient links). The update rule for a node representation on the $k^{th}$ encoder layer is defined as: 

\begin{equation} \begin{split}
h_{v}^{(k)}= \sigma\left( W_{1}^{(k)} \cdot h_{v}^{(k-1)} + W_{2}^{(k)} \cdot 
\mathrm{MEAN}(\{h_{u}^{(k-1)} \mid u\in N(v)\})\right), \\
\end{split}\end{equation}
where $h_{v}^{(k)}$ is the representation of node v at the $k$ layer and  $N(v)$ denotes the set of neighbours of node $v$, $\mathrm{MEAN}$ refers to the average operation. $W_{1}^{(k)}$ and $W_{2}^{(k)}$ are two learnable weight matrix. Notably, we use the original features from each omic as the input to the first GNN layer. We set the number of layers to be 2 for each GNN encoder. Lastly, the shared projection layers comprise stacked MLP layers which ingest the node representations from the final GNN layer and output the final patient embedding $e_{v}$:

\begin{equation} \begin{split}
e_{v}= \mathrm{MLP}(h_{v}^{(N)}). \\
\end{split}\end{equation}

\paragraph{Learning objective}

For better illustration, again consider the integration of two distinct data modalities: $\mathbf{X^{(1)}} \in \mathbb{R}^{n_{x1} \times d_{x1}} $ and $\mathbf{X^{(2)}} \in \mathbb{R}^{n_{x2} \times d_{x2}} $.  In the IntegrAO embedding phase, our objective is to map these datasets into a unified embedding space of dimensionality $q$. The resultant lower-dimensional datasets are represented as $\mathbf{X^{(1)'}} \in \mathbb{R}^{n_{x1} \times q }$ and $\mathbf{X^{(2)'}} \in \mathbb{R}^{n_{x2} \times q}$. This embedding is achieved by optimizing two distinct loss functions: the reconstruction loss $L_{reconc}$ and the alignment loss $L_{align}$. The reconstruction loss $L_{reconc}$ is conceptualized on the principles of t-distribution stochastic neighbor embedding (t-SNE)\cite{van2008visualizing} and can be formally defined as:

\begin{equation} \begin{split}
L_{\mathrm{reconc}} = \mathrm{KL}(P^{(1)} \mid\mid Q^{(1)}) + \mathrm{KL}(P^{(2)} \mid\mid Q^{(2)}),
\end{split}\end{equation}
where $P^{(1)}$ and $P^{(2)}$ are the fused patient graphs obtained during the fusion stage with diagonal values set to 0. And $Q^{(1)}$ and $Q^{(2)}$, constrained to t-distribution, are the sample-to-sample transition probability matrix calculated using the low-dimensional embedding $\mathbf{X^{(1)'}}$ and $\mathbf{X^{(2)'}}$. The Kullback-Leibler (KL) divergences is defined as:

\begin{equation} \begin{split}
 \mathrm{KL}(P \mid\mid Q) = \sum_{i}\sum_{j}P_{ij}\mathrm{log}\frac{P_{ij}}{Q_{ij}}.
\end{split}\end{equation}
The alignment loss $L_{align}$ quantifies the mean squared error between embeddings of the same patients derived from different omics modalities. It is defined as:

\begin{equation} \begin{split}
L_{\mathrm{align}} = \frac{1}{n} \sum_{i=1}^{c} \mathbbm{1}(i \in \mathbf{C})  (X^{(1)'}_{i} - X^{(2)'}_{i})^{2},
\end{split}\end{equation}
where $C$ denotes the set of common samples between the two modalities. The final loss is the combination of reconstruction loss and alignment loss as:

\begin{equation} \begin{split}
Loss = L_{\mathrm{reconc}}  + \beta \times L_{\mathrm{align}},
\end{split}\end{equation}
where $\beta$ is a tradeoff parameter to balance the KL terms and the embedding alignment term. We set $\beta$=1 in all our experiments. The model can be readily extended to multi-view data by adding additional KL divergence terms to the reconstruction loss for each added view, and summing all pairwise alignment losses between modalities for the matching loss.  We solve the optimization problem using gradient descent with a fixed number of epochs. We set epoch=1000 in all our experiments. 
\paragraph{Model output}

After training, the final output is derived by averaging patient embeddings across modalities. Let $M(i)$ denote the omic types available for patient $i$, then the final patient embeddings $E(i)$ for patient $i$ are obtained by:

\begin{equation} \begin{split}
E(i) = \frac{1}{\mid M(i) \mid}\sum_{m \in M(i)} e_{i}^{(m)},
\end{split}\end{equation}
where $e_{i}^{(m)}$ is the embedding for patient $i$ from modality $m$. The final integrated network is then computed using Eq. \ref{matrix_sim} followed by Eq. \ref{stable_norm}, taking the final patient embeddings as input.

\subsection{Inductive Prediction}\label{fine-tune}

\paragraph{Model fine-tuning for subtype prediction}

After unsupervised integration of multi-omics data, patient subtypes can be determined by clustering the final integrated network. Given defined subtype labels, IntegrAO can be further fine-tuned to predict subtypes for new patients based on any combination of omics data. To enable this, we initialized with the unsupervised IntegrAO model parameters and appended a prediction head that ingests the final patient embeddings to output a subtype prediction. We calculate the classification loss $L_{\mathrm{clf}}$ as:

\begin{equation} 
\begin{split}
L_{\mathrm{clf}} = -\frac{1}{N} \sum_{i=1}^{N} \left( y_i \log(\mathrm{Pred}(E_{i})) + (1 - y_i)\log(1 - \mathrm{Pred}(E_{i})) \right) , 
\end{split}
\end{equation}
where $\mathrm{Pred(\cdot)}$ is the fully connected prediction head, and $y_i$ are the defined subtype labels. During fine-tuning, we jointly optimize the total loss:

\begin{equation} \begin{split}
Loss = L_{\mathrm{reconc}}  + \beta \times L_{\mathrm{align}} + \gamma \times L_{\mathrm{clf}}.
\end{split}\end{equation}
The hyperparameter $\beta$ and $\gamma$ control the tradeoff between the reconstruction loss, alignment loss, and classification loss during optimization.

\paragraph{Subtype prediction for new patient}\label{model inference}

During supervised fine-tuning, for the omics data used in training, let $\{\mathbf{X_{tr}}^{(m)} \mid m= 1, 2, ..., v \}$ denote the input omic features for different modalities, and $\{\mathbf{P_{tr}}^{(m)} \mid m= 1, 2, ..., v \}$ the corresponding fused similarity matrix.  The fine-tuned IntegrAO model can then be trained on $\{\mathbf{X_{tr}}^{(m)}\}$ and $\{\mathbf{P_{tr}}^{(m)}\}$, with training predictions represented as: 

\begin{equation} \begin{split}
\mathbf{Y_{tr}} = \mathrm{IntegrAO}(\{\mathbf{X_{tr}}^{(m)}\}, \{\mathbf{P_{tr}}^{(m)}\}), m= 1, 2, ..., v,
\end{split}\end{equation}
where $\mathbf{Y_{tr}} \in \mathbb{R}^{n_{tr} \times c}$ contains the predicted subtype probabilities for each of the $n_{tr}$ training samples, with $c$ denoting the number of subtypes. For a new test sample 
$\{\mathbf{X_{te}}^{(m)} \mid m= 1, 2, ..., v \}$, to perform model inference, we extend the data matrix of the corresponding omics to $ \{ \mathbf{X_{trte}} = 
\begin{bmatrix}
X_{tr} \\
X_{te}
\end{bmatrix} \mid m= 1, 2, ..., v \}
$, and generate the extended fusion matrix by performing the fusion step with the testing samples $\{\mathbf{P_{trte}}^{(m)} \mid m= 1, 2, ..., v \}$. Therefore, given $\{ \mathbf{X_{trte}} \}$ , $\{ \mathbf{P_{trte}} \}$ and fine-tuned $\mathrm{IntegrAO}$ model, we have: 

\begin{equation} \begin{split}
\mathbf{Y_{trte}} = \mathrm{IntegrAO}(\{\mathbf{X_{trte}}^{(m)}\}, \{\mathbf{P_{trte}}^{(m)}\}), m= 1, 2, ..., v,
\end{split}\end{equation}
where $\mathbf{Y_{trte}} \in \mathbb{R}^{n_{tr+1} \times c}$.
The predicted subtype probability distribution for the testing sample is at the last row of $\mathbf{Y_{trte}}$. 

\subsection{Cluster number selection for AML subtyping and cancer patient classification expeiments}\label{choose number}

To identify the optimal number of clusters for cancer datasets, we implemented a specific approach. First, after integrating patient data, we conducted a 10-fold train-test split. In each fold, a Gaussian Mixture Model (GMM) was applied to 90\% of the patient embeddings, and log-likelihood scores were calculated on the remaining 10\%. This process was repeated for cluster numbers in a pre-defined range. We then computed the mean and standard deviation of the log-likelihood scores for each cluster number. The optimal cluster number was determined by the log-likelihood score, calculated by subtracting the mean from the standard deviation, and used to rank the suitability of each cluster number for the dataset.

In the new patient classification experiments, spectral clustering with this optimal cluster number was applied to the integrated network to obtain clustering labels. For the AML case study, an initial identification of 18 clusters was refined by merging biologically similar clusters, resulting in 12 distinct AML subtypes.

\subsection{Gene expression deconvolution}\label{cell deconvolution}

To generate the cell composition data for our cancer benchmarking experiments, we utilized \textit{BayesPrism}\cite{chu2022cell} to deconvolute raw gene expression counts from TCGA cancer cohorts. Our analyses were conducted exclusively through the \textit{BayesPrism} web portal, adhering to its default preprocessing steps. These steps included filtering outlier genes, selecting protein-coding genes, and isolating signature genes for each cell type. For deconvolution job submissions, we employed the portal's default settings. The resulting matrices, detailing fractions of patient-specific cell types, served as the cell composition modality for our integration benchmarking. The single-cell reference datasets utilized in the deconvolution process are detailed in \textbf{Supplementary Table S4}.

\backmatter

\section*{Declarations}

\bmhead{Code Availability}

The code to utilize IntegrAO is available on Github: \url{https://github.com/bowang-lab/IntegrAO}.

\newpage





\bibliography{bib_v1}

\newpage

\section{Supplementary Tables}

\setcounter{table}{0}
\makeatletter 
\renewcommand{\thetable}{S\@arabic\c@table}
\makeatother

\begin{table}[h!]
    \centering
    \caption{The number of patients used in the benchmarking analysis per cancer type. It specifies the patient counts across six diagnostic modalities: mRNA expression, cellular composition, DNA methylation, miRNA expression, reverse-phase protein array, and copy-number variation. The table further details the intersection and comprehensive aggregates of patients within each cancer category.}
    \begin{tabular}{|c|c|c|c|c|c|c|c|c|}
        \hline
        Dataset & \multicolumn{1}{p{0.8 cm}|}{mRNA} & \multicolumn{1}{p{1.3 cm}|}{Cell Composition} & \multicolumn{1}{p{0.8cm}|}{Meth} & \multicolumn{1}{p{0.8cm}|}{miRNA} & \multicolumn{1}{p{0.8cm}|}{RPPA} & \multicolumn{1}{p{0.8cm}|}{CNV} & \multicolumn{1}{p{1.0cm}|}{Common patients} & \multicolumn{1}{p{1.0cm}|}{Union patients}  \\
        \hline
        \hline
        BRCA & 1093 & 1093 & 1080 & 756 & 784 & 887 & \underline{511} & \underline{1096} \\
        COAD & 379 & 379 & 616 & 295 & 393 & 494 & \underline{251} & \underline{621} \\
        SKCM & 469 & 469 & 367 & 448 & 470 & 353 & \underline{247} & \underline{461}  \\
        KIRC & 533 & 533 & 528 & 257 & 319 & 478 & \underline{306} & \underline{537}  \\
        LUAD & 515 & 515 & 516 & 457 & 458 & 365 & \underline{166} & \underline{511}  \\
        \hline
    \end{tabular}
    \label{tab: cancer size}
\end{table}

\newpage

\begin{table}[h!]
    \centering
    \caption{The number of features used for each omic in the benchmarking analysis per cancer type. The six modalities including: mRNA expression, cellular composition, DNA methylation, miRNA expression, reverse-phase protein array, and copy-number variation.}
    \begin{tabular}{|c|c|c|c|c|c|c|}
        \hline
        Dataset & \multicolumn{1}{p{0.8 cm}|}{mRNA} & \multicolumn{1}{p{1.3 cm}|}{Cell Composition} & \multicolumn{1}{p{0.8cm}|}{Meth} & \multicolumn{1}{p{0.8cm}|}{miRNA} & \multicolumn{1}{p{0.8cm}|}{RPPA} & \multicolumn{1}{p{0.8cm}|}{CNV}  \\
        \hline
        \hline
        BRCA & 2000 & 25 & 2000 & 897 & 222 & 2000  \\
        COAD & 2000 & 8 & 2000 & 623 & 222 & 2000  \\
        SKCM & 2000 & 9 & 2000 & 901 & 195 & 2000   \\
        KIRC & 2000 & 13 & 2000 & 825 & 212 & 2000   \\
        LUAD & 2000 & 8 & 2000 & 894 & 195 & 2000   \\
        \hline
    \end{tabular}
    \label{tab: feature dim}
\end{table}

\newpage

\begin{table}[h!]
    \centering
    \caption{Number of clusters chosen for the new patient classification experiment. Spectral clustering with this chosen cluster count was performed on the IntegrAO-integrated network to determine the ``ground truth'' labels for each cancer type.}
    \begin{tabular}{c | c c c c c }
        \hline
         & \multicolumn{1}{p{0.8 cm}}{BRCA} & \multicolumn{1}{p{0.8 cm}}{COAD} & \multicolumn{1}{p{0.8cm}}{SKCM} & \multicolumn{1}{p{0.8cm}}{KIRC} & \multicolumn{1}{p{0.8cm}}{LUAD} \\
        \hline
        Number of cluster & 5 & 7 & 5 & 5 & 3 \\

        \hline
    \end{tabular}
    \label{tab: cluster num}
\end{table}

\newpage

\begin{table}[h!]
    \centering
    \caption{ Number of cells and cell types in the single-cell reference data used for gene expression deconvolution across TCGA cancers. This cell type count includes various subtypes of tumor cells.}
    \begin{tabular}{|c|c|c|c|}
        \hline
        Cancer type & \multicolumn{1}{p{1.3 cm}|}{Cell number} & \multicolumn{1}{p{1.3 cm}|}{Cell type number} & \multicolumn{1}{p{1.3 cm}|}{Reference} \\
        \hline
        \hline
        BRCA & 45561 & 70 & \cite{azizi2018single}   \\
        COAD & 18409 & 14 & \cite{lee2020lineage}  \\
        SKCM & 6879 & 19 & \cite{jerby2018cancer}    \\
        KIRC & 20476 & 24 & \cite{li2022mapping}    \\
        LUAD & 52698 & 8 & \cite{lambrechts2018phenotype}   \\
        \hline
    \end{tabular}
    \label{tab: cell number}
\end{table}

\newpage

\setcounter{figure}{0}
\makeatletter 
\renewcommand{\thefigure}{S\@arabic\c@figure}
\makeatother

\section{Supplementary Figures}



\begin{figure}[h!]
    \centering
    \includegraphics[width=11cm]{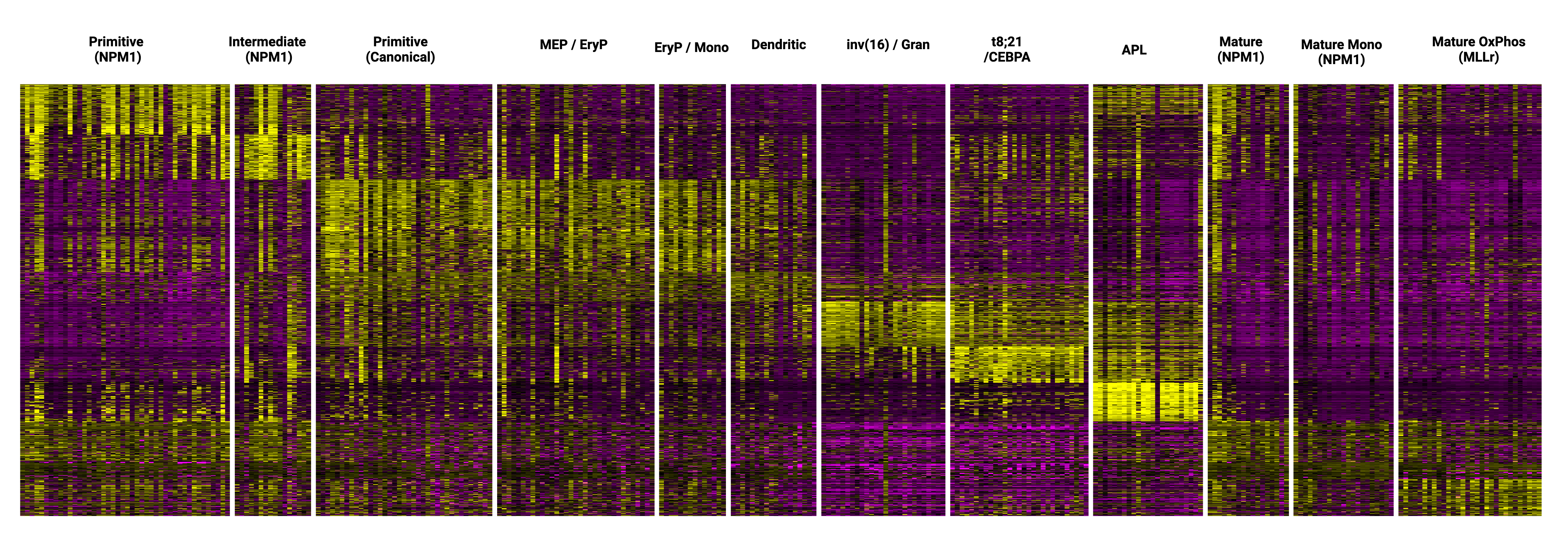}
    \caption{IntegrAO identification of 12 subtypes with distinct DNA methylation profiles. The visual representation shows a clear block structure, effectively delineating each cluster, highlighting the distinctiveness of methylation patterns among the subtypes.}
    \label{aml_meth}
\end{figure}
\newpage

\begin{figure}[h!]
    \centering
    \includegraphics[width=11cm]{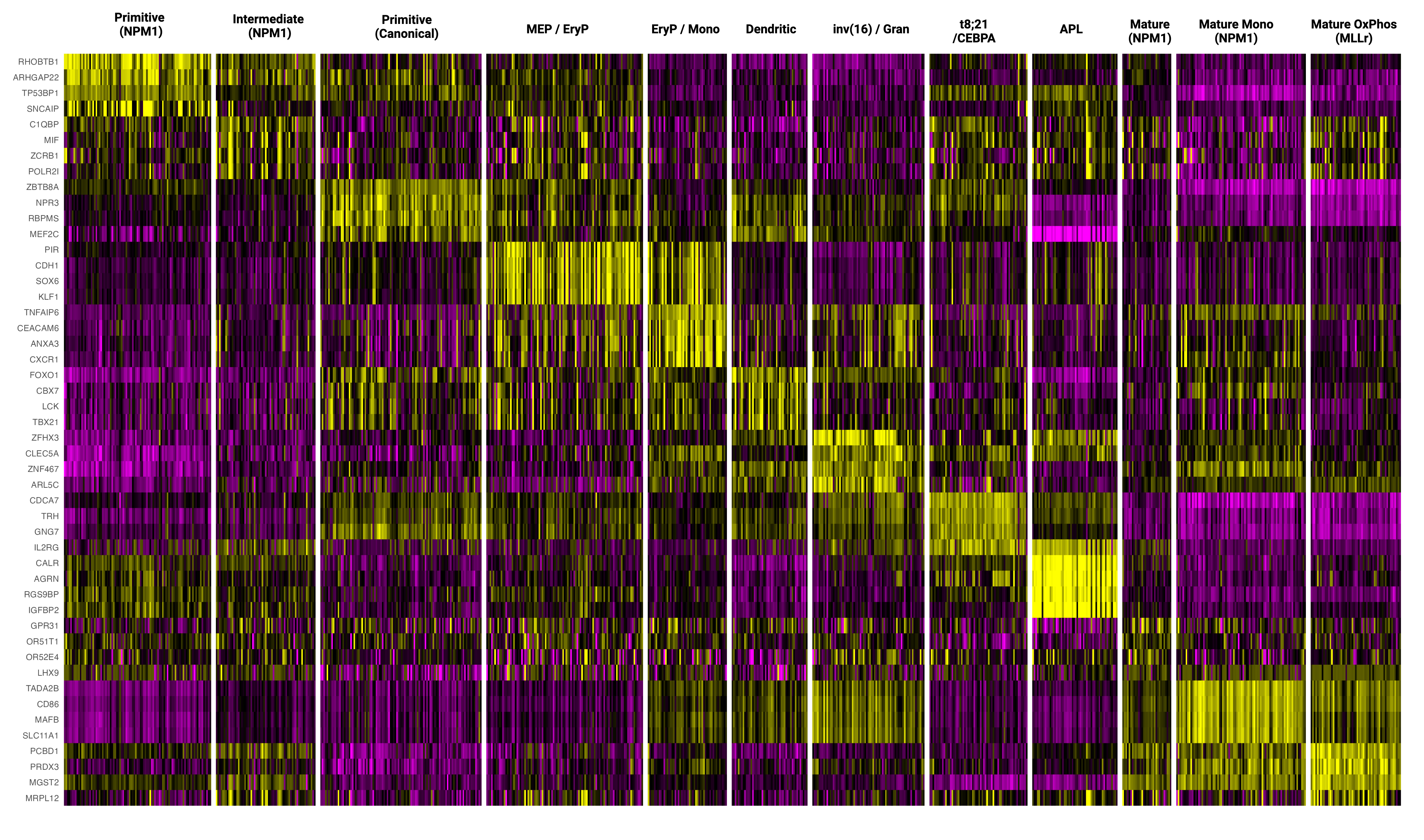}
    \caption{VIPER analysis of all regulons: heatmap showcasing cluster-specific regulatory signatures. This heatmap reveals distinct block structures corresponding to IntegrAO-defined clusters, illustrating the diverse regulatory landscapes encompassing all types of regulons, including transcription factors, non-coding RNAs, and other regulatory molecules.}
    \label{aml_all_regulator}
\end{figure}
\newpage

\begin{figure}[h!]
    \centering
    \includegraphics[width=11cm]{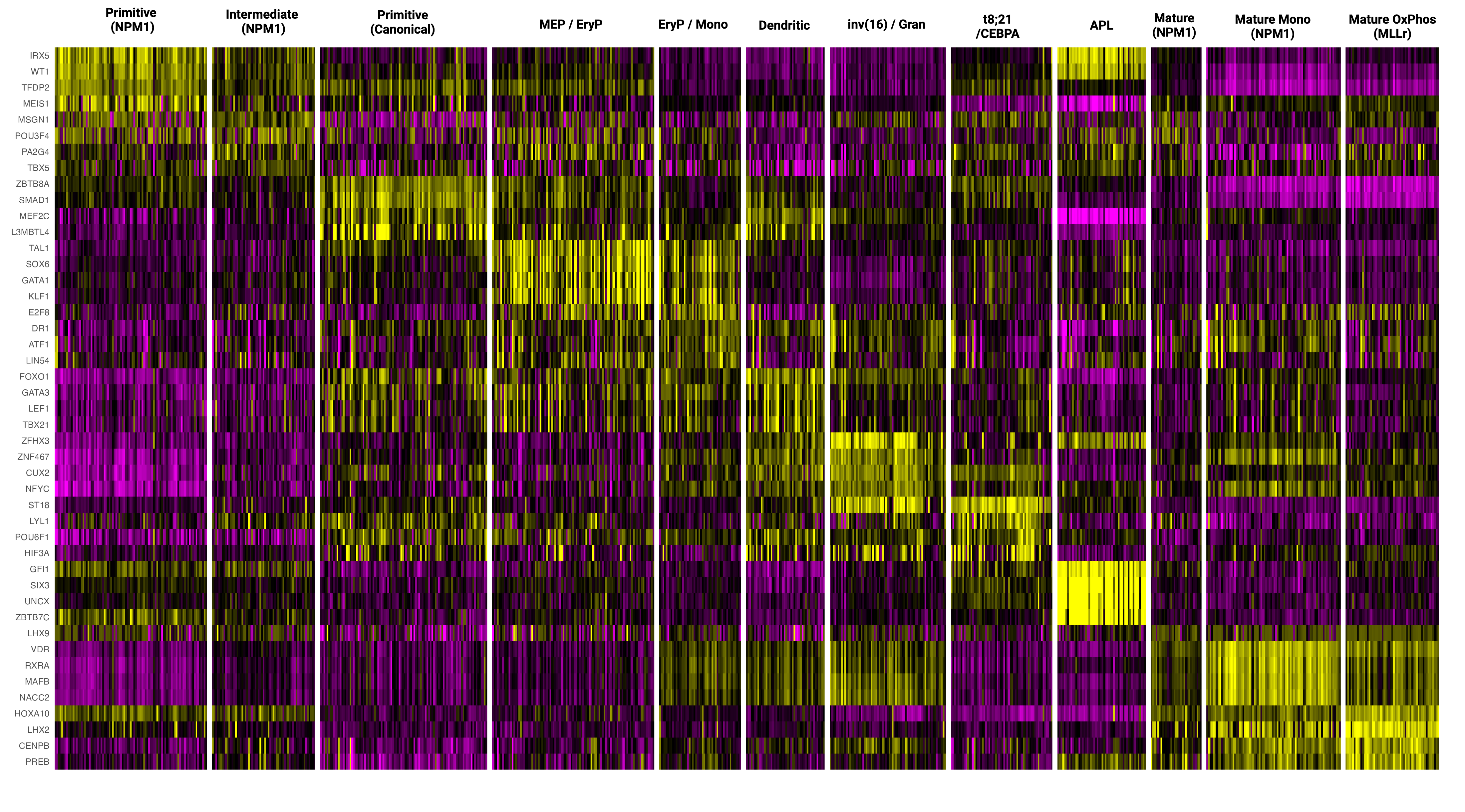}
    \caption{VIPER analysis of transcription factor (TF) regulons: heatmap depicting transcription factor-driven regulatory patterns. The heatmap displays clear, distinct blocks that align with IntegrAO-defined clusters, highlighting the specific influence of transcription factors on the gene expression within each cluster.}
    \label{aml_tf_regulator}
\end{figure}
\newpage

\begin{figure}[h!]
    \centering
    \includegraphics[width=11cm]{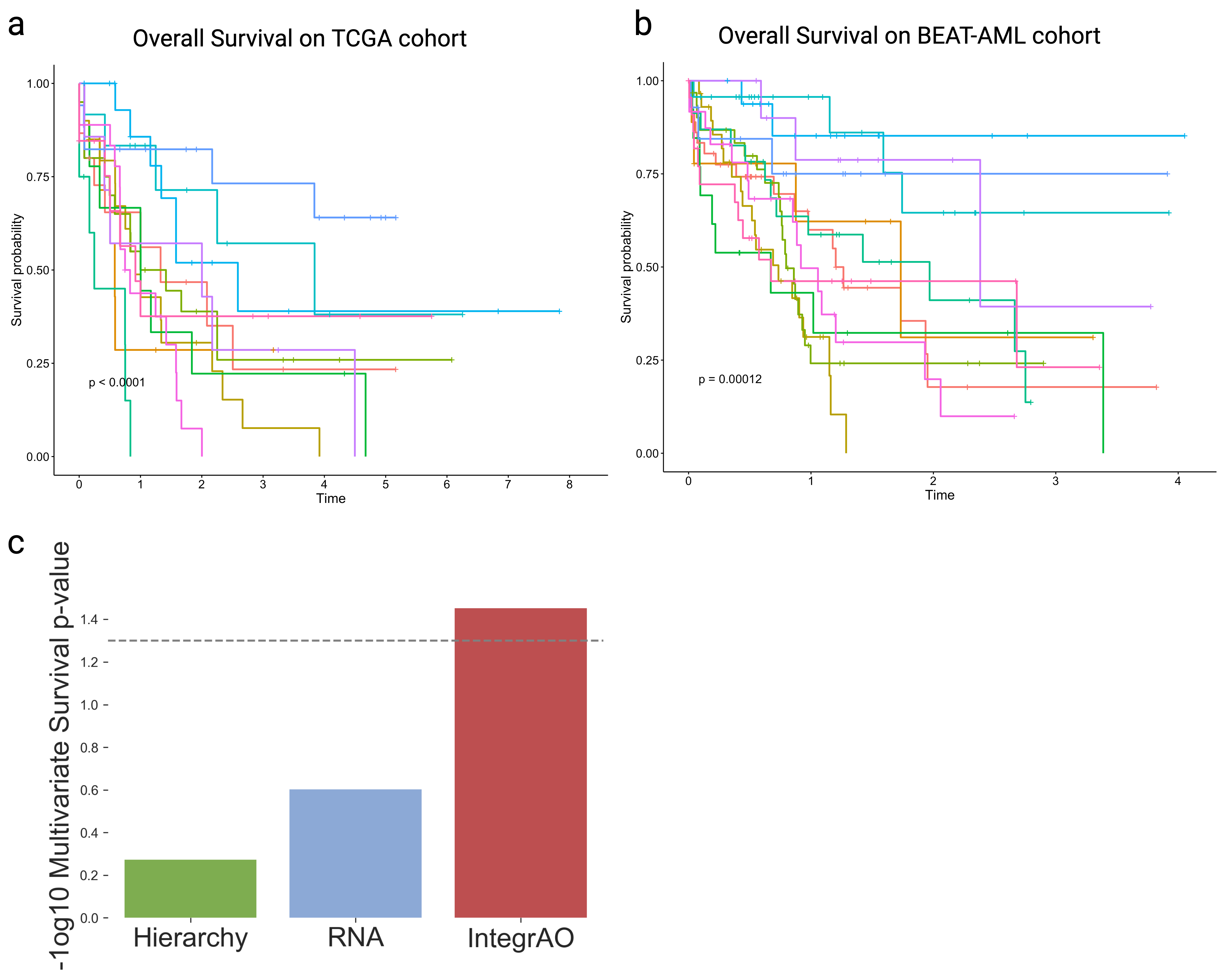}
    \caption{Additional survival analysis of the AML case study includes: (a) a Kaplan-Meier survival curve for the TCGA AML patient cohort, stratified by IntegrAO's clusters, showing statistical significance (multi-group logrank test p-value = 6.1e-5); (b) a similar Kaplan-Meier curve for the BEAT-AML patient cohort, also stratified by IntegrAO's clusters, indicating significant differences in survival outcomes (multi-group logrank test p-value = 1.2e-4); (c) a comparison of multivariate survival significance between IntegrAO's clustering solution and solutions derived from using only cell hierarchy or RNA data on the TCGA and BEAT-AML combined cohort. This comparison demonstrates that IntegrAO's clustering notably enhances multivariate survival significance.}
    \label{aml_survival_curves}
\end{figure}
\newpage

\begin{figure}[h!]
    \centering
    \includegraphics[width=11cm]{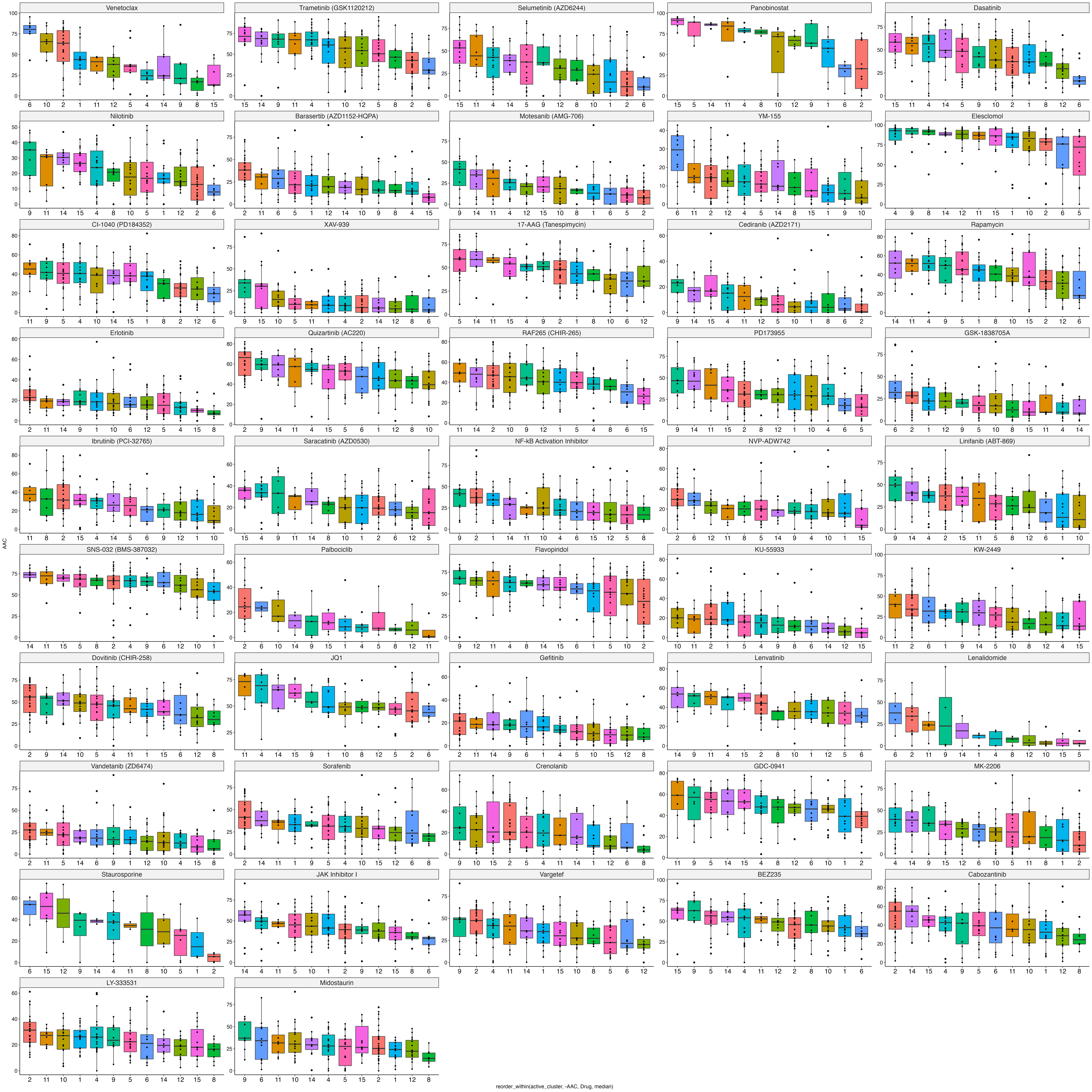}
    \caption{Drug sensitivity profile across IntegrAO-identified clusters. This figure illustrates the differential responses of various IntegrAO-derived clusters to a range of anti-cancer agents, showcasing the distinct drug sensitivity patterns characteristic of each cluster in the context of AML treatment.}
    \label{aml_drugs}
\end{figure}
\newpage

\begin{figure}[h!]
    \centering
    \includegraphics[width=11cm]{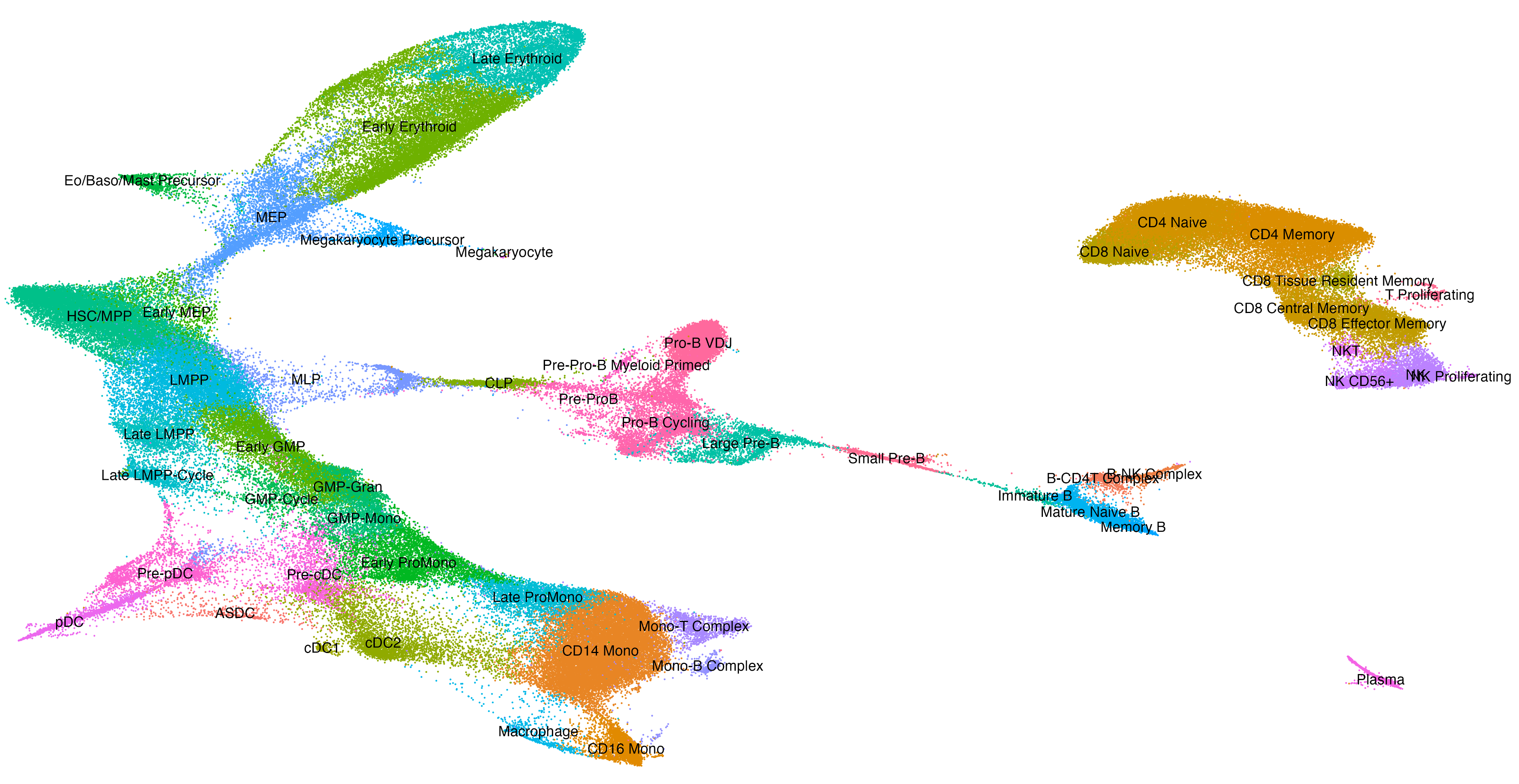}
    \caption{UMAP Visualization of single-cell RNA-seq data from bone marrow mononuclear cells, based on research by \textit{Galen et al}\cite{van2019single}. This plot offers a detailed representation of cell-type diversity and distribution within the bone marrow environment, as captured through advanced single-cell sequencing techniques.}
    \label{scRNA_AML}
\end{figure}
\newpage

\begin{figure}[h!]
    \centering
    \includegraphics[width=11cm]{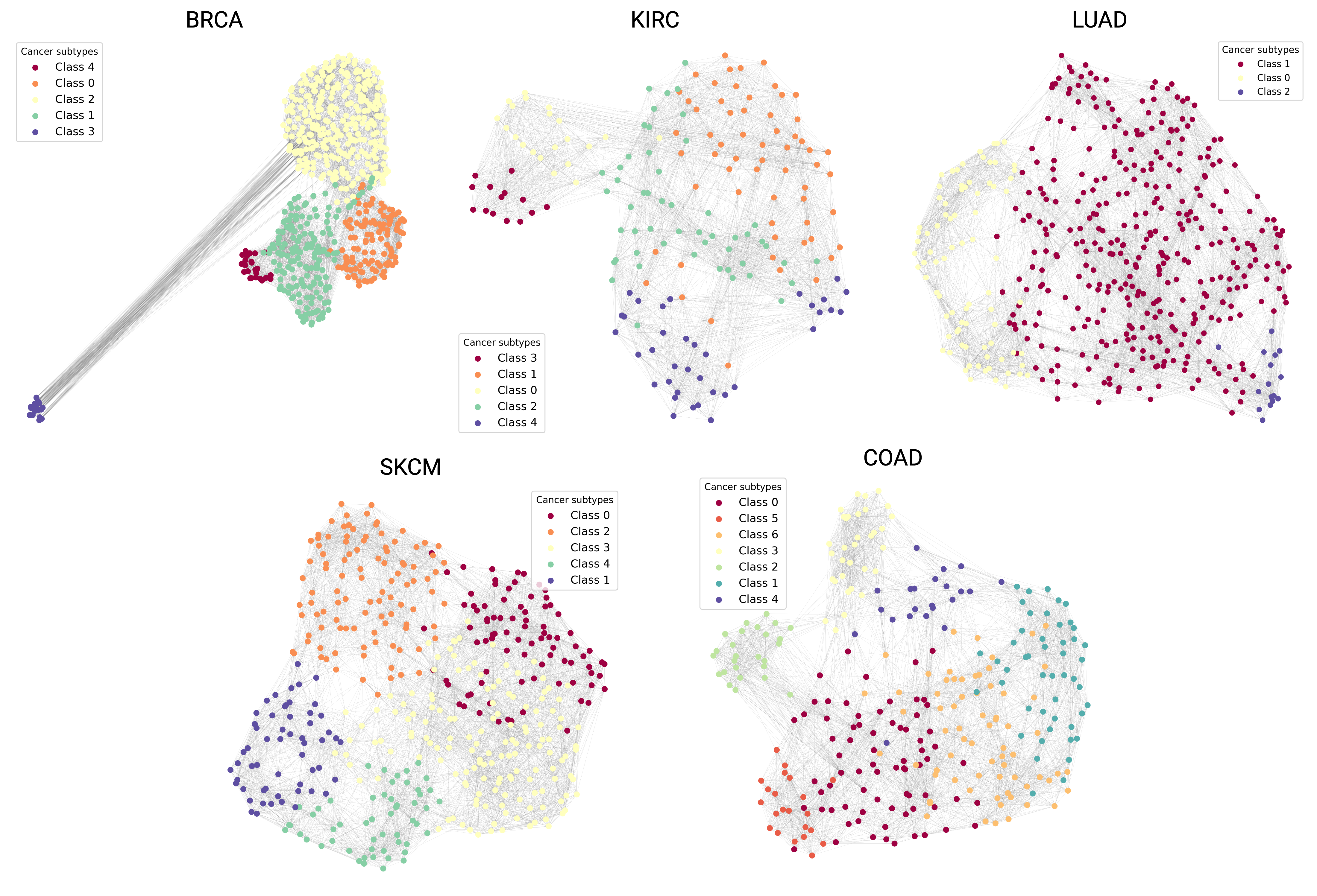}
    \caption{UMAP plots displaying patient embeddings across five cancer types from the new patient classification experiments. Labels were derived via spectral clustering on the full integrated network, using the preselected number of clusters.}
    \label{cancer_projection_umap}
\end{figure}
\newpage

\end{document}